\documentclass[
    10pt
    ,aps
    ,pra
    ,oneside
    ,nobibnotes
    ,floatfix
    ,twocolumn
    ,nofootinbib
    ,superscriptaddress
    ,showkeys
]{revtex4-2}

\usepackage{amsmath,amssymb,bm}
\usepackage{array}
\allowdisplaybreaks[3]
\usepackage{siunitx}
\NewDocumentCommand{\bnum}{O{}m}{%
  \num[math-rm=\mathbf,color=Blue,#1]{#2}%
}

\usepackage{graphicx}\graphicspath{{}{Graph/}{Graph/seminar/}{../Graph/}{../../Graph/}}
\usepackage[svgnames]{xcolor}

\usepackage{iftex}
\iftutex
    \usepackage{fontspec}

    \usepackage[math-style=ISO,bold-style=ISO]{unicode-math}
\else

    \usepackage[T1]{fontenc} 
    \usepackage[utf8]{inputenc} 
\fi

\usepackage[english, showlanguages]{babel}

    \iftutex
        
    \else
        
    \fi

    \DeclareMathOperator{\const}{const}

    \DeclareMathOperator{\K}{K}

    \newcommand{\dif}[2][]{\mathop{}\!\mathrm{d}
        \if
            \relax\detokenize{#1}\relax
        \else
            ^{\mkern-1.mu#1}\mkern-2.5mu 
    \fi
    #2\,}
    \newcommand{\der}[2]{\frac{\dif{#1}}{\dif{#2}}}
    \newcommand{\tder}[2]{{\dif{#1}}/{\dif{#2}}}

    \newcommand{\mean}[1]{\left\langle #1\right\rangle}

\usepackage[
    colorlinks
        ,linkcolor=violet
        ,filecolor=purple
        ,citecolor=teal
        ,urlcolor=magenta
    ,unicode
    ,pdfpagemode=UseOutlines
    ,pdfdisplaydoctitle=true
    ,pdfpagetransition=Wipe
    ,pdfusetitle
]{hyperref}
\hypersetup{
        pdfkeywords={ballooning modes, MHD stability, Gas-Dynamic Trap}
}
\usepackage{bookmark}

\begin{document}

\def\bot{\mathrel\perp}

\title{Wall stabilization of the rigid ballooning $m=1$ mode in a long-thin mirror trap}

\author{Igor Kotelnikov}
    \email{I.A.Kotelnikov@inp.nsk.su}
    \affiliation{Budker Institute of Nuclear Physics SB RAS, Novosibirsk, 630090, Russia}
\author{Qiusun Zeng}
    \email{qiusun.zeng@inest.cas.cn}
    \affiliation{Institute of Nuclear Energy Safety Technology HFIPS CAS, Hefei, 230031, People’s Republic of China}

\author{Vadim Prikhodko}
    \email{V.V.Prikhodko@inp.nsk.su}
\author{Dmitri Yakovlev}
    \email{D.V.Yakovlev@inp.nsk.su}
    \affiliation{Budker Institute of Nuclear Physics SB RAS, Novosibirsk, 630090, Russia}

%
\author{Keqing Zhang}
    \email{keqingz@mail.ustc.edu.cn}
\author{Zhibin Chen}
    \email{zhibin.chen@inest.cas.cn}
\author{Jie Yu}
    \email{yujie@inest.cas.cn}
    \affiliation{Institute of Nuclear Energy Safety Technology HFIPS CAS, Hefei, 230031, People’s Republic of China}
\date{\today}

\begin{abstract}
    The prospect of stabilization of the $m=1$ ``rigid'' ballooning mode in an open axially symmetric long-thin trap with the help of a conducting lateral wall surrounding a column of isotropic plasma is studied. It is found that for effective wall stabilization, the beta parameter must exceed $70\%$. The dependence of the critical beta on the mirror ratio, the radial pressure profile, and the axial profile of the vacuum magnet has been studied. It is shown that when a conductive lateral wall is combined with conductive end plates simulating attachment of the end MHD stabilizers to the central cell of an open trap, there are two critical beta values and two stability zones that can merge, making stable the entire range of allowable beta values $0<\beta<1$.
\end{abstract}
\maketitle

\section{Introduction}\label{s1}

%
Continuing the study of ballooning instability started in the article \cite{Kotelnikov+2021PST_24_015102}, in this paper we present the results of calculating the critical beta ($\beta_{}$, the ratio of the plasma pressure to the magnetic field pressure) for rigid ballooning perturbations with azimuth number $m=1$ in a mirror trap (also called open trap). A proper ballooning equation for a plasma with diffuse plasma pressure radial profile was derived by Lynda LoDestro \cite{LoDestro1986PF_29_2329} in 1986, but, in fact, neither she nor anyone else ever used it. Probably oblivion for many years of LoDestro's work is due to the early termination of the TMX (Tandem Mirror eXperiment) and MFTF-B (Mirror Fusion Test Facility B) projects in the USA in the same 1986 \cite{Ongena+2016NaturePhysics_12_398}. However, achievement of a high electron temperature and high beta in the Gas-Dynamic Trap (GDT) at the Budker Institute of Nuclear Physics in Novosibirsk
\cite{
    Ivanov+2003PhysRevLett_90_105002, 
    Simonen+2010JFE_29_558, 
    Bagryansky+2011FST_59_31,
    Bagryansky+2015PhysRevLett_114_205001,
    Bagryansky+2015NF_55_053009,
    Bagryansky+2016AIPConfProc_030015,
    Bagryansky+2016AIPCP_1771_020003,
    Yakovlev+2018NF_58_094001,
    Bagryansky+2019JFE_38_162},
as well as emergence of new ideas \cite{Beklemishev2016PoP_23_082506} and new projects \cite{Granetzny+2018APS_CP11_150, Bagryansky+2020NuclFusion_60_036005} makes us rethink old results.

Unlike many previous works
\cite{
    KaiserNevinsPearlstein1983PF_26_351, 
        Berk+1985PPCNFR_2_321,
        Berk+1984PF_27_2705, 
        HaasWesson1967PF_10_2245, 
        DIppolitoHafizi1981PF_24_2274, 
        DIppolitoMyra1984PF_27_2256,
    KaiserPearlstein1985PhysFluids_28_1003,
    Kesner1985NF_25_275,
    LiKesnerLane1985NF_25_907, 
            LiKesnerLane1987NF_27_101,
        LiKesnerLane1985NF_25_907, 
        LiKesnerLoDestro1987NF_27_1259
},
%
whose authors studied the stability of the rigid ballooning mode in a plasma model with a radial pressure profile shaped as a step or a ring with sharp boundaries (sharp-boundary or staircase models),  LoDestro derived an equation for a plasma with a diffuse radial profile of pressure in the paraxial (also called long-thin) approximation. It is suitable for describing both isotropic and anisotropic plasmas.

%

In this paper, the LoDestro equation is used to calculate the critical value of the parameter beta, above which an isotropic plasma will be stabilized by a lateral perfectly conductive wall surrounding the plasma column. The calculations are made for four different radial plasma profiles and many axial profiles of the vacuum magnetic field, some of which we previously used, studying the stability of small-scale ballooning perturbations \cite{Kotelnikov+2021PST_24_015102}.
In addition, we study the effect of wall stabilization in combination with the action of conductive end plates, which imitate stabilization by the end magnetohydrodynamic (MHD) anchors.

%
The isotropic plasma approximation does not quite adequately describe the plasma state in open traps, except for traps with a very large mirror ratio. Calculations show that the critical beta in a plasma with an anisotropic pressure can be much smaller than that found by the isotropic plasma approximation. In other words, wall stabilization of an anisotropic plasma is more efficient than wall stabilization of an isotropic plasma. Our study of ballooning instability in an anisotropic plasma will be reported in a subsequent paper. Numerical solution of the LoDestro equation in the case of an anisotropic plasma is much more complicated and time-consuming. This circumstance, as well as the abundance of new results even for the case of isotropic plasma, motivated our decision to separate the case of isotropic plasma into a distinct article in order to describe the calculation technique in more detail here.

%
To avoid possible misunderstanding (and objections), it should be clarified that ballooning modes are usually understood as $z$-dependent pressure-gradient-driven modes with $m\gg1$. Some authors object to the use of the term ``ballooning'' for the $m=1$ mode, in which the internal deformations in the cross section of the plasma column are frozen due to the effects of Finite Larmor Radius (FLR effects). These authors propose to call such modes ``rigid'' or ``global''. Nevertheless, referring to the $m=1$ rigid mode as a ballooning mode did not stop, therefore we prefer to call the oscillations studied in this paper as ``rigid ballooning mode'', although such a name does not seem quite satisfactory to us.


In what follows, we will adhere to the following plan of presentation. In the next section \ref{s2} a review of the publications on the stability of the $m=1$ ballooning mode that preceded LoDestro's paper is given; after this article, publications on ballooning instability in open traps practically ceased. In  section \ref{s3}, the LoDestro equation is written and the necessary notation is introduced. Section \ref{s4} presents the results of calculating the critical beta in the limit when a perfectly conducting wall surrounding the plasma column almost closely adjoins the lateral boundary of the column, but does not touch it. In this limit, the LoDestro ordinary differential equation reduces to an integral over the $z$ coordinate along the trap axis; the integral vanishes at the critical value of beta. Section \ref{s5} describes the solution of the LoDestro equation by the shooting method and presents the results of calculations for several model pressure and magnetic field profiles. In section \ref{s6}, the shooting method is again used to solve the LoDestro equation with different boundary conditions that model the effect of conductive end plates placed in magnetic mirrors. Final section \ref{s9} summarizes our results and conclusions.

\section{Literature review}\label{s2}

There are a number of publications, in one way or another, related to the stability of ballooning MHD perturbations in mirror traps. Most of them were published in the 1980s. In the next decades, interest in the problem of ballooning instability in mirror traps significantly weakened (in contrast to what is happening in tokamaks, see e.g.\ \cite{Snyder+2002PoP_9_2037, Halpern+2013PoP_20_052306, Eich+2018NF_58_034001}), which was a consequence of the termination of the TMX
and MFTF-B
projects in the USA in 1986 \cite{Ongena+2016NaturePhysics_12_398} as mentioned in Section \ref{s1}.
%
%
A review of publications devoted to the stability of small-scale ballooning oscillations with a large azimuth number $m\gg1$ was made in our recent paper \cite{Kotelnikov+2021PST_24_015102}. We will not repeat it here and immediately turn to works on the stability of ballooning oscillations with the azimuthal number $m=1$, which are ``rigid'' in a certain sense as explained below.

%

1.
According to modern views, small-scale flute and ballooning oscillations with a large azimuthal number $m\gg 1$ must be stabilized due to the effects of a finite Larmor radius. This conclusion follows from the fundamental work of \emph{Rosenbluth, Krall and Rostoker} \cite{Rosenbluth+1962NFSuppl_1_143}, where the role of FLR effects is revealed using the kinetic equation; see also \cite{RobertsTaylor1962PhysRevLett_8_197, Rudakov1962NF_2_107}, where it is proved that the FLR effects can be included into equations of magnetohydrodynamics if the viscous stress tensor is preserved. In paraxial open traps, FLR effects can in principle stabilize all modes, except for oscillations with an azimuthal number $m=1$. A more accurate estimate of the number of azimuthal modes that the effects of FLR stabilize is obtained in the article \cite{DIppolitoFrancisMyraTang1981PF_24_2270}. The FLR effects impose on oscillations with azimuthal number $m=1$ the form of a rigid ("solid-state") displacement without deformation of the plasma interior in each cross section. However, even in this case, the displacement of the plasma column from the axis varies in different sections. As a result, the plasma column is bent. The bend is most noticeable in the region of the so-called unfavorable curvature near the minimum value of the magnetic field in the central section of the axially symmetric open trap. It is these (balloon) oscillations that we study in this article.


2.
\emph{Kaiser, Nevins, and Pearlstein} in a 1983 paper \cite{KaiserNevinsPearlstein1983PF_26_351} investigated the stability of the $m=1$ rigid mode in a quadrupole open trap in the paraxial approximation for a low-pressure plasma at $\beta\to0$. These authors did not assume  presence of a conductive wall around the plasma, but actually took into account the effects of FLR, since they considered only rigid displacements of the plasma.

3.
\emph{Berk et al} in 1984 showed with a kinetic treatment that a perfectly conducting wall located near the lateral plasma surface {in case of large beta} has a strong stabilization effect on the $m = 1$ mode in an axisymmetric mirror, which cannot otherwise be stabilized by FLR effects \cite{Berk+1985PPCNFR_2_321, Berk+1984PF_27_2705}.
%
%
These authors argued (in our opinion, not quite objectively) that in previous works the effect of wall stabilization was overlooked. They wrote:
    ``Previous analyses either did not take boundary conditions into account properly \cite{KaiserNevinsPearlstein1983PF_26_351} or were for isotropic pressure \cite{HaasWesson1967PF_10_2245,DIppolitoHafizi1981PF_24_2274}, where beta of order unity is needed for stability.''
Berk et al analyzed the effect of wall stabilization and fast electron ring on  curvature-driven modes, drift modes, anisotropy-driven modes such as AIC, however ballooning modes are not explicitly mentioned. The source of stability is the image currents generated by placing the wall (or properly shaped conductors) in close proximity of a high-beta axially localized plasma.

4.
%
The MHD approach to the study of wall stabilization historically preceded Berk's theory. Later it was inherited and supplemented by several authors
\cite{
    HaasWesson1967PF_10_2245, 
    DIppolitoHafizi1981PF_24_2274, 
    DIppolitoMyra1984PF_27_2256,
    KaiserPearlstein1985PhysFluids_28_1003,%
    Kesner1985NF_25_275%
}.
For simplicity, Refs.~\cite{
    Berk+1984PF_27_2705,
    HaasWesson1967PF_10_2245,
    DIppolitoHafizi1981PF_24_2274,
    KaiserPearlstein1985PhysFluids_28_1003
    }
assumed a sharp boundary pressure profile.
%
%
In particular, \emph{Haas and Wesson} in 1967 considered the hydromagnetic stability of a theta-pinch with a sharp boundary \cite{HaasWesson1967PF_10_2245}. They allowed for the magnetic field to vary along the pinch axis, so that, in fact, they analyzed stability of a mirror trap.
They found that the \emph{necessary} condition for stability of the $m=1$ mode  is $\beta  > \beta_{\text{crit}} = 1/[1 + (a/r_{w})^{2}]$, where $a$ and $r_{w}$ are the radii of the plasma and conducting wall. In other words, $\beta > 50\%$ is required in the limit $a\to r_{w}$ in agreement with later publications.


5.
\emph{D'Ippolito and Hafizi} in 1981 employed a simplified  model of an axisymmetric tandem mirror to study the stability of low-$m$ ballooning modes in isotropic plasma with a sharp boundary surrounded by a perfectly conducting wall \cite{DIppolitoHafizi1981PF_24_2274}. The scaling of the critical beta $\beta_{\text{crit}}$ with mode number $m$, connection length $L_{c}$ to the end plug, and the wall to plasma radius ratio $r_{w}/a$ was computed for several magnetic field profiles. Important simplification was that the end plugs of the tandem mirror are not explicitly modeled but are replaced by a boundary condition on the perturbation, viz., that the field lines are “tied” at some distance $L_{c}$ outside the central cell. The authors found a second zone of stability at large beta, but in the case of isotropic plasma they considered, the second zone appeared only when beta was very close to unity.

6.
\emph{D'Ippolito and Myra} in 1984 numerically analyzed stability of the $m=1$ rigid ballooning mode in an axisymmetric tandem mirror with inverted pressure profile \cite{DIppolitoMyra1984PF_27_2256}. Included in the analysis are the stabilizing effects of an externally applied force, such as the rf-induced ponderomotive force, and of a perfectly conducting lateral wall. The authors assumed isotropic plasma with a hollow stepwise pressure profile and studied wall stabilization. They found two zones of stability at low and high betas. These two zones merge in case when the conducting wall is located sufficiently close to the plasma lateral boundary and the radial pressure profile has the shape of thin annular.


7.
\emph{Kaiser and Pearlstein} in their 1985's paper \cite{KaiserPearlstein1985PhysFluids_28_1003} wrote equations for studying the stability of the $m=1$ mode in an axially symmetric trap. They took into account the FLR effects and conductive wall in a plasma model with an arbitrary beta, but with a stepwise radial profile. These authors start by writing out Eq.~((1))\footnote{
    We use double brackets to denote numbers of equations in cited papers.
}, which they say can be ``synthesized'' from three papers \cite{PearlsteinFreidberg1978PF_21_1278, Newcomb1981JPP_26_529, Newcomb1973AnnPhys_81_231}. This equation contains a term that takes into account the FLR effects.
%
%
Equation ((1)) is then used to derive equation ((11)), which was later used by Li, Kesner and Lane to analyze the stability of the rigid mode in \cite{LiKesnerLane1985NF_25_907}, where it appears under the number ((28)). For the case when the conducting wall of the vacuum chamber is located as close as possible to the lateral surface of the plasma, equation ((11)) is simplified to equation ((19)), which was later studied by other authors, including Kesner in \cite{Kesner1985NF_25_275}.

8.
In the same year 1985, \emph{Kesner} in Ref.~\cite{Kesner1985NF_25_275} discussed the possibility of an axisymmetric tandem mirror in which stability accrues from wall stabilization. The author used an anisotropic plasma model with a sharp boundary with reference to the above-cited work by Kaiser and Pearlstein \cite{KaiserPearlstein1985PhysFluids_28_1003}, and he studied the case when the chamber walls were located extremely close to the lateral surface of the plasma. In this case, as shown in \cite{KaiserPearlstein1985PhysFluids_28_1003}, the ballooning mode becomes almost fluted. It was also assumed that the vacuum magnetic field has a parabolic profile up to magnetic mirrors. The axial distribution of the plasma pressure was given by two versions of the function $p_{\perp}(B)$. One distribution corresponded to the pressure maximum in the median plane of the trap, the other distribution described the plasma with sloshing ions, when the pressure maximum was reached in the gap between the median plane and the magnetic mirrors. The author begins his analysis with equation ((1)), which is identical to equation ((19)) in paper \cite{KaiserPearlstein1985PhysFluids_28_1003} by Kaiser and Pearlstein.

%
%
It was shown that plasma stability is achieved if the parameter $\beta_{}$ exceeds a certain limiting value $\beta_{\text{crit}}$, which depends on the degree of plasma anisotropy: the smaller is the limiting value, the stronger is the anisotropy. For sufficiently large anisotropy, $\beta_{\text{crit}}$ decreases to $0.4$. For an isotropic plasma, this value increases to $0.8$.


%

9.
In paper \cite{LiKesnerLane1985NF_25_907}, also published in 1985, \emph{Li, Kesner and Lane} used the MHD energy principle to examine the stabilization effect of a conducting wall located near the plasma lateral surface. It was assumed that conducting wall is the only stabilization mechanism.


The calculation starts with equation ((28)), which coincides with equation ((11)) from the cited above article \cite{KaiserPearlstein1985PhysFluids_28_1003}  by Kaiser and Pearlstein. In contrast to their own equation ((19)) in equation ((11)) the assumption has not yet been made that the conducting wall is located close to the lateral surface of the plasma, that is, the parameter $\Lambda =(r_{w}^{2}+ a^{2})/(r_{w}^{2}-a^{2})$ is not equal to infinity. However, then the authors only analyze the limit when $\Lambda=\infty$. In this case, the plasma displacement $\xi_{n}$ turns out to be quasiflute, i.e.\ $aB_{v}\xi_{n}\approx\const$ (where $B_{v}$ is the magnetic field in the vacuum gap), and the parameter $\Lambda $ is knocked out of the equation by integrating it with respect to the variable $z$ along the axis of the system under the condition that at the ends of the integration interval the displacement is not frozen into the ends (free-end boundary condition), i.e. $(aB_{v}\xi_{n})'=0$ (where the prime $'$ stands for derivative over coordinate $z$ along the trap axis), while, as the authors prove, normal to boundary of the magnetic field perturbation component $\delta B_{n}$ vanishes.


Although the intermediate formulas are written for an anisotropic plasma, the final analysis is limited to the case of an isotropic plasma. For an isotropic pressure component, it is found that a hollow profile has better stability than a uniform pressure when the integral of the radial pressure profile is fixed.

10.
\emph{Same authors} in a latter paper \cite{LiKesnerLane1987NF_27_101}, published in 1987, discussed the wall stabilization by partially enclosed wall using $m=1$ model and stepwise radial profile of isotropic plasma.
%
%
The stabilizing wall extends axially only over a part of the distance between the trap midplane and the mirror throat. The wall is located near the plasma surface in the bad curvature region and far from it in the good curvature region. A variational method is used to solve the equations for both regions, with the authors solving equation ((1)), which is the same as equation ((13)) from paper \cite{KaiserPearlstein1985PhysFluids_28_1003} cited above. At the ends of the plasma column, the boundary condition $(B_{v}\xi_{n}/\sqrt{B})'=0$ was used, which for a plasma with a sharp boundary is equivalent to the boundary condition $(aB_{v}\xi_{n})'=0$.
For the connection of the regions of close and distant plasma-wall proximity, a jump condition is used. The variational calculation is performed with a simple trial function (the choice of the trial function is substantiated with an exact numerical solution). The results show that (i)~the removal of the conducting wall in the good curvature region does not significantly degrade plasma stability, (ii)~the acceptable ratio $r_{w}/a$ of the radius of the conducting wall $r_{w}$ to the plasma radius $a$ is about $1.1$, and (iii)~for cases with a low mirror ratio, more conducting wall is needed for stability than for cases with a high mirror ratio.

%
11.
In the next paper \cite{LiKesnerLoDestro1987NF_27_1259} of 1987, \emph{Li, Kesner and LoDestro} have shown that a simple axisymmetric magnetic mirror may be MHD stable, provided that (i)~a certain length of magnetic field has a series of ripples in it, (ii)~with isotropic pressure the critical beta is higher than 50\%, and (iii)~the conducting wall is very close to the plasma surface. The theory of ballooning instability and its physical picture are discussed, and a Sturm-Liouville form is presented as well as numerical results that highlight the requirements of the field structure and plasma anisotropy. The authors numerically solve the equation ((1)), which is derived under the number ((13)) in the paper \cite{KaiserPearlstein1985PhysFluids_28_1003} for an anisotropic plasma with sharp boundary. The calculation is performed both for isotropic plasma and for an anisotropic plasma, in which the transverse pressure varies in magnetic field according to the law $p_{\bot}\propto B_{\max}^{2}-B^{2}$ or $p_{\bot}\propto (B/B_{\max})^{2}(1-B/B_{\max})^{n-1}$. It is shown that anisotropy reduces the beta required for stability, particularly at low mirror ratio. Apparently, this is the first work in which the critical beta is calculated at a finite value of $\Lambda$, i.e. at a nonzero width of the vacuum gap between the plasma and the side conducting wall of the chamber.

%
12.
\emph{LoDestro} in the paper \cite{LoDestro1986PF_29_2329} of 1986 derived a ballooning equation for the $m = 1$ mode in an axisymmetric plasma with a diffuse pressure profile within the framework of the arbitrary-$\beta_{}$, dominant FLR analysis of Kaiser and Pearlstein \cite{KaiserPearlstein1985PhysFluids_28_1003}. According to the author, diffuse profiles preserve the sharp-boundary result that only the vacuum curvature appears in the destabilizing term. It is shown that the diffuse profile reduces the volume-averaged pressure, which is necessary for the stability of an isotropic plasma in the limit when the conducting wall approaches the plasma/vacuum interface and the mirror ratio approaches unity.


13.
\emph{Close and Lichtenberg} in their 1989 paper \cite{CloseLichtenberg1989PFB_1_629} reported the results of an experiment on the MMX device at Berkeley. High-beta ballooning modes are studied in an axisymmetric multiple mirror which is made average-minimum $B$ with end cusps. Electric and magnetic field measurements in the plasma characterize the predominant azimuthal mode number as $m = 1$. The ballooning character of the mode is determined by measuring the ratio of the mode amplitude near the device center to that near the cusp, and confirmed by measurement of perturbed perpendicular magnetic fields. Theoretical growth rates are calculated numerically using ideal and resistive magnetohydrodynamic equations for the rigid $m = 1$ ballooning mode. Within experimental error it is found that the $m = 1$ resistive ballooning growth rate scales with radially averaged beta $\langle\beta \rangle$ approximately as $\langle\beta \rangle^{1/2}$ for $\langle\beta \rangle\lesssim 0.10$ (on axis $\beta_{}\lesssim 0.20$), in agreement with theory. The observed growth rates increase with mirror ratio as expected. The resistive growth rates calculated numerically agree reasonably well with experimental observations.
%
%
Details of the theoretical calculations are not given, but it is indicated that numerical codes were used.

14.
\emph{Kang, Lichtenberg, and Nevins} in the paper \cite{KangLichtenbergNevins1987PF_30_1416} of 1987 have developed ideal and resistive MHD computer codes for the rigid $m = 1$ ballooning mode in the Berkeley Multiple Mirror. The numerical codes are based on the fluid equation of Lee and Catto \cite{LeeCatto1981PF_24_2010} and include the effects of nonparaxial curvature in the cusp. These codes are used to calculate theoretical growth rates for comparison with experiment. The authors note that the calculation without taking into account the resistivity gives the limiting beta larger than was found from the measurement results.

\section{LoDestro equation}\label{s3}

In its final form, the LoDestro equation has the form
    \begin{multline}
    \label{3:01}
    0 = \der{}{z}
    \left[
        \Lambda + 1 - \frac{[4\pi]\mean{p_{\bot}+p_{\|}}}{B_{v}^{2}}
    \right]
    \der{\phi}{z}
    \\
    +
    \phi
    \left[
        - \der{}{z}\left(
            \frac{B_{v}'}{B_{v}} + \frac{2a'}{a}
        \right)
        \left(
            1 - \frac{[4\pi]\mean{p_{\bot}+p_{\|}}}{2B_{v}^{2}}
        \right)
    \right.
    \\
    \left.
    +
    \frac{\omega^{2}\mean{\rho}}{B_{v}^{2}}
    -
    \frac{[4\pi]\mean{p_{\bot}+p_{\|}}}{B_{v}^{2}}\frac{a_{v}''}{a_{v}}
    \right.
    \\
    \left.
    -
    \frac{1}{2}\left(
            \frac{B_{v}'}{B_{v}} + \frac{2a'}{a}
    \right)^{2}
    \left(
        1 - \frac{[4\pi]\mean{p_{\bot}+p_{\|}}}{2B_{v}^{2}}
    \right)
    \right]
    ,
    \end{multline}
where the derivative $\tder{}{z}$ in the first two lines acts on all factors to the right of it, and the prime ``$'$'' is a shortcut for $\tder{}{z}$. The required function
    \begin{equation}
    \label{3:02}
    \phi(z) = a(z) B_{v}(z) \xi_{n}(z)
    /
    \sqrt{2\psi_{a}}
    \end{equation}
%
depends on one coordinate $z$ along the trap axis and is expressed in terms of the plasma boundary radius $a=a(z)$, the vacuum magnetic field $B_{v}=B_{v}(z)$, and the small displacement $\xi_{n}=\xi_{n}(z)$ of the plasma column from the axis. The normalizing denominator $\sqrt{2\psi_{a}}$ in Eq.~\eqref{3:02} is present in LoDestro's article, but we omit it below. It doesn't make much sense, since $\psi_{a}$ is a constant, $\psi_{a}=\const$. The parameter $\psi_{a}$ has the meaning of the reduced (i.e.\ divided by $2\pi$) magnetic flux through the plasma cross section $\pi a^{2}$. It is related to the plasma radius $a=a(z)$ by the equation
    \begin{equation}
    \label{3:03}
    \frac{a^{2}}{2} = \int_{0}^{\psi_{a}} \frac{\dif{\psi}}{B}
    .
    \end{equation}
The magnetic flux $\psi$ through an arbitrary circular section of the plasma and the radial coordinate $r$ relates the equation
    \begin{equation}
    \label{3:03a}
    \frac{r^{2}}{2} = \int_{0}^{\psi} \frac{\dif{\psi}}{B}
    .
    \end{equation}
%
The magnetic field $B=B(\psi,z)$ in the paraxial (long-thin) approximation (i.e.\ with a small curvature of field lines) is related to the vacuum magnetic field $B_{v}=B_{v}(z)$ by the transverse equilibrium equation
    \begin{equation}
    \label{3:04}
        B^{2} = B_{v}^{2} -[4\pi]\,2p_{\bot}
    .
    \end{equation}
%
The factor $[4\pi]$ in Eqs.~\eqref{3:01} and~\eqref{3:04} arises in the Gaussian system of units. LoDestro and some other authors use rationalized electromagnetic units (also known as Heaviside—Lorentz units) where the factor $[4\pi]$ is dropped. In what follows, we also omit it.

The kinetic theory predicts (see, for example, \cite{Newcomb1981JPP_26_529}) that the longitudinal and transverse plasma pressures can be considered as functions of $B$ and $\psi$, i.e. $p_{\bot}=p_{\bot}(B ,\psi)$, $p_{\|}=p_{\|}(B,\psi)$. In Eq.~\eqref{3:01}, one must assume that the magnetic field $B$ is already expressed in terms of $\psi$ and $z$, and therefore $p_{\bot}=p_{\bot}(\psi,z)$, $p_{\|}=p_{\|}(\psi,z)$. In what follows, we will also use the notation
    \begin{equation}
    \label{3:05}
    \overline{p} = \frac{p_{\bot} + p_{\|}}{2}
    .
    \end{equation}
%
One should distinguish between the actual plasma radius $a=r(\psi_{a},z)$ and the vacuum plasma radius
    \begin{equation}
    \label{3:06}
    a_{v}(z) = \sqrt{\frac{2\psi_{a}}{B_{v}(z)}}
    .
    \end{equation}
%
It enters Eq.~\eqref{3:01} as the ratio $a_{v}''/a_{v}$, where the prime denotes the derivative of $\tder{}{z}$ with respect to the coordinate $z$. LoDestro draws the reader's attention to the fact that only the vacuum field line curvature $a_{v}''$ enters into the equation, but in fact the curvature $a''$ of the plasma boundary arises when calculating the derivative in the second line of the equation. We also point out that $\rho$ is the plasma mass density, and $\omega $ is the oscillation frequency.

%
The angle brackets in Eq.~\eqref{3:01} denote the average
    \begin{gather}
    \label{3:07}
    \mean{g}
    =
    \frac{
        \int_{0}^{\psi_{a}} \dif{\psi} g/B
    }{
        \int_{0}^{\psi_{a}} \dif{\psi}/B
    }
    =
    \frac{2}{a^{2}}
    \int_{0}^{\psi_{a}} \frac{\dif{\psi}}{B}\,g
    \end{gather}
of an arbitrary function $g(\psi,z)$ over the plasma cross section.
Parameter
    \begin{equation}
    \label{3:08}
        \Lambda = \frac
    {
        r_{w}^{2} + a^{2}
    }{
        r_{w}^{2} - a^{2}
    }
    \end{equation}
%
is expressed in terms of the actual radius of the plasma/vacuum boundary $a=a(z)$ and the radius of the conducting cylinder $r_{w}=r_{w}(z)$, which surrounds the plasma column. The parameter $\Lambda =\Lambda (z)$ is generally a variable function of the $z$ coordinate, but in the remainder of the paper we assume that $\Lambda $ is a constant. The larger the $\Lambda $ value, the closer the conducting cylinder is to the plasma boundary. The $\Lambda \to \infty $ limit corresponds to the case when the conducting side wall is close to the plasma boundary, repeating its shape, but does not touch the plasma. The limit $\Lambda \to 1$ means that the lateral conducting wall is removed to infinity.

%
We have repeated the derivation of the LoDestro equation and now we are sure that it is correct, although there are typos in a pair of intermediate formulas in Ref.~\cite{LoDestro1986PF_29_2329}.

%
The boundary conditions for Eq.~\eqref{3:01} and similar equations in the study of ballooning instability are traditionally set at the ends of the plasma column at the magnetic field maxima $B_{v}=B_{m}$, where $B_{v}'=0$ and $p_{\bot}=p_{\|}=0$. In accordance with the geometry of actually existing open traps, it is usually assumed that the magnetic field is symmetrical with respect to the median $z=0$ plane, and the magnetic mirrors (i.e., field maxima) are located at $z=\pm L$.

Traditionally, two types of boundary conditions are considered. In the presence of conductive end plates directly in magnetic mirrors, it is required that the boundary condition be satisfied
    \begin{equation}
    \label{3:11}
    \phi = 0
    \end{equation}
%
at $z=\pm L$. A similar boundary condition is usually used in studying the stability of small-scale ballooning disturbances, thereby modeling the presence of a stabilizing cell behind a magnetic mirror (see, for example, \cite{Kotelnikov+2021PST_24_015102}).

If the plasma ends are isolated, the boundary condition
    \begin{equation}
    \label{3:12}
    \phi' = 0
    \end{equation}
is applied.
%
As a rule, it implies that other methods of MHD stabilization in addition to stabilization by a conducting lateral wall are not used. It is this boundary condition \eqref{3:12} that was used earlier in the works on the stability of the $m=1$ ballooning mode.


\section{Limit of zero vacuum gap}
\label{s4}

For this section, Eq.~\eqref{3:01} is reduced for the limit $\Lambda \to \infty $  as the lateral conducting wall approaches the plasma/vacuum boundary, where it produces its maximum stabilizing effect. In this limit it is also possible to make analytic progress in assessing the effects of a diffuse profile.

%
Stabilization of the rigid ballooning mode by a conducting wall in the $\Lambda \to \infty $ limit was previously studied by \emph{Kesner} \cite{Kesner1985NF_25_275}, \emph{Li, Kesner and Lane} \cite{LiKesnerLane1985NF_25_907} in the case of a plasma with a sharp-boundary radial profile. They showed that the plasma displacement $\xi_{n}$ in this limit turns out to be quasiflute, i.e.\ $\phi=aB_{v}\xi_{n}\approx\const$, and the parameter $\Lambda $ is knocked out from an equation like Eq.~\eqref{3:01} by integrating it over the coordinate $z$ along the axis of the mirror trap, if the insulating boundary condition \eqref{3:12} is allowed.
For $\Lambda \to \infty$ the first term in Eq.~\eqref{3:01} is formally greater than all the others, so the derivative $\tder{\phi}{z}$ must tend to zero in proportion to $1/\Lambda$, i.e.
    \begin{equation}
    \label{4:01}
    \phi = \phi_{0} + \delta\phi(z)
    ,
    \end{equation}
%
moreover, $\delta\phi(z)=\mathcal{O}(1/\Lambda)$, and the constant $\phi_{0}$ can be considered equal to $1$ due to the linearity of Eq.~\eqref{3:01} with respect to the function $\phi$. Substituting $\phi=\phi_{0}=1$ into the second term in Eq.~\eqref{3:01} (that's the whole square bracket on three lines) and integrating the whole equation from $z=-L$ to $z=L$ drops out the first term (with large $\Lambda $) provided that the boundary condition $\phi'=\delta\phi'=0$ is used at $z=\pm L$.
Performing the indicated procedure yields  the integral equation
    \begin{multline}
    \label{4:02}
    \omega^{2}
    \int_{-L}^{L}
    \frac{\mean{\rho}}{B_{v}^{2}}
    \dif{z}
    =
    \left.
    \left(
        \frac{B_{v}'}{B_{v}} + \frac{2a'}{a}
    \right)
    \left(
        1 - \frac{\mean{\overline{p}}}{B_{v}^{2}}
    \right)
    \right|_{-L}^{+L}
    +\\+
    \int_{-L}^{L}
    \left[
        \frac{2\mean{\overline{p}}}{B_{v}^{2}}\frac{a_{v}''}{a_{v}}
        +
        \frac{1}{2}\left(
            \frac{B_{v}'}{B_{v}} + \frac{2a'}{a}
        \right)^{2}
        \left(
            1 - \frac{\mean{\overline{p}}}{B_{v}^{2}}
        \right)
    \right]
    \dif{z}
    ,
    \end{multline}
%
where the factors $[4\pi]$ before $\mean{\overline{p}}=\mean{p_{\bot}+p_{\|}}/2$ are omitted for brevity. It allows one to calculate the squared oscillation frequency $\omega^{2}$ if the radial profile of pressure $\overline{p}$, density $\rho$, and vacuum magnetic field $B_{v}$ are known. At the margins of the stable regime, the oscillation frequency is equal to zero, $\omega^{2}=0$. This fact is proved in the theory of ideal magnetohydrodynamics (see, for example, \cite{Bateman1978mhd, Freidberg1987, Kotelnikov2021V1e}). In the stability region $\omega^{2}>0$, and instability takes place if $\omega^{2}<0$.


%
The bracket $\left( {B_{v}'}/{B_{v}} + {2a'}/{a} \right)$ is proportional to $B_{v}'$, so the first term on the right-hand-side of Eq.~\eqref{4:02} is zero if (as we assume) the boundary conditions at $z=\pm L$ are set in the throats of magnetic mirrors, where $B_{v}'=0$.

In the case of a plasma with a sharp boundary, we have
    \begin{multline}
    \label{4:03}
    \frac{B_{v}'}{B_{v}} + \frac{2a'}{a}
    =
    \frac{B_{v}'}{B_{v}} - \frac{B'}{B}
    =
    -\frac{(B/B_{v})'}{B/B_{v}}
    =\\=
    - \frac{1}{2}\frac{(1-\beta_{\bot})'}{1-\beta_{\bot}}
    =
    \frac{1}{2}\frac{\beta_{\bot}'}{1-\beta_{\bot}}
    ,
    \end{multline}
%
where $B=B_{v}\sqrt{1-\beta_{\bot}}$, $a^{2}=2/B$, $\beta_{\bot} = 2p_{\bot}/B_{ v}^{2}$, so Eq.~\eqref{4:02} becomes
    \begin{multline}
    \label{4:04}
    \omega ^{2} \int_{-L}^{L} \frac{\rho}{B_{v}^{2}} \dif{z}
    =\\=
    \int_{-L}^{L} \left[
        \frac{2\overline{p}}{B_{v}^{2}}\frac{a_{v}''}{a_{v}}
        +
        \frac{1}{8}\left(
            \frac{\beta_{\bot}'}{1-\beta_{\bot}}
        \right)^{2}
        \left(
            1 - \frac{\overline{p}}{B_{v}^{2}}
        \right)
    \right]
    \dif{z}
    .
    \end{multline}
%
This result coincides with equation ((19)) in Kaiser and Pearlstein's paper \cite{KaiserPearlstein1985PhysFluids_28_1003}. The first term in square brackets on the right-hand-side is a generalization of the Rosenbluth–Longmire criteria \cite{RosenbluthLongmire1957AnnPhys_1_120}; in the case of an isotropic plasma, the integral of this term is always negative (see, for example, \cite{Kotelnikov2021V2e}). The second term is certainly positive, but for $\beta \ll 1$ it is less than the first one. This means that $\omega^{2}<0$ in the limit of $\beta\to0$. Therefore, if stability is possible, then only if beta exceeds some limiting value, $\beta > \beta_{\text{crit}} > 0$.

%

%
Further in this section, we present the results of calculating the critical value of beta,  $\beta_{\text{crit}}$, which corresponds to the marginal stability $\omega ^{2}=0$ in an isotropic plasma in the limit $\Lambda = \infty $. It is known that the pressure $p=p_{\bot}=p_{\|}$ in an isotropic plasma is constant along the field line (see, for example, \cite{Freidberg1987, Bateman1978mhd, Kotelnikov2021V2e}), i.e. $\overline{p}=p$ is actually independent of $B$ and therefore also of $z$, so we can write
    \begin{equation}
    \label{4:05}
    p=p_{0}f_{k}(\psi)
    ,
    \end{equation}
%
where $p_{0}=\const$ is the plasma pressure on the trap axis, and the dimensionless function $f_{k}(\psi)$ is defined so that $f_{k}(0)=1$.

%
In the calculations below, we used dimensionless variables, denoting them in the same way as their dimensional counterparts. Dimensionlessness is achieved by taking the values of the constants $L=1$ and $\psi_{a}=1$. We normalize the magnetic field by the value of the function $B_{v}(z)$ at $z=0$, so that for the dimensionless function $B_{v}(0)=1$. Let us also define the dimensionless parameter
    \begin{equation}
    \beta_{}=2p_{0}/B_{v}^{2}(0)=2p_{0}
    ,
    \end{equation}
so that
    \begin{gather}
    \label{4:06}
    B=\sqrt{B_{v}^{2}-\beta_{}f_{k}}
    ,\\
    \label{4:07}
    \frac{a^{2}}{2} = \int_{0}^{1} \frac{\dif{\psi}}{B}
    ,\\
    \label{4:08}
    \frac{a_{v}^{2}}{2} = \frac{1}{B_{v}}
    ,\\
    \label{4:09}
    \mean{\overline{p}}
    =
    \frac{\beta_{}}{a^{2}}\int_{0}^{1} \frac{f_{k}}{B}\dif{\psi}
    .
    \end{gather}
%
The critical value of the parameter $\beta_{}$ corresponding to the marginal stability $\omega ^{2}=0$ is defined as the root of the equation
    \begin{equation}
    \label{4:10}
    W(\beta_{\text{crit}}) = 0
    ,
    \end{equation}
where
    \begin{equation}
    \label{4:11}
    W(\beta_{})
    =
    \int_{-1}^{1}
    \left[
        \frac{2\mean{\overline{p}}}{B_{v}^{2}}\frac{a_{v}''}{a_{v}}
        +
        \frac{1}{2}\left(
            \frac{B_{v}'}{B_{v}} + \frac{2a'}{a}
        \right)^{2}
        \left(
            1 - \frac{\mean{\overline{p}}}{B_{v}^{2}}
        \right)
    \right]
    \dif{z}
    .
    \end{equation}

%
%

\begin{figure*}
  \centering
  \includegraphics[width=0.3\linewidth]{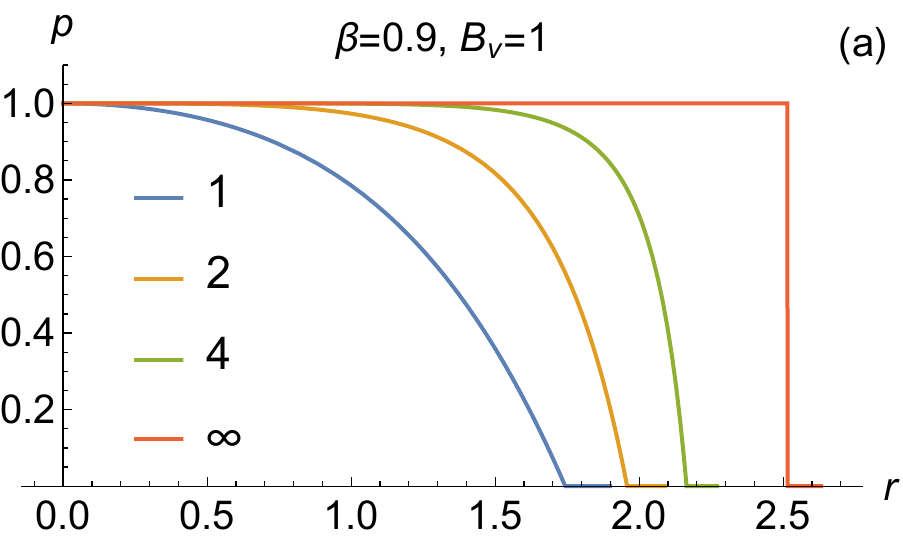}\hfil
  \includegraphics[width=0.3\linewidth]{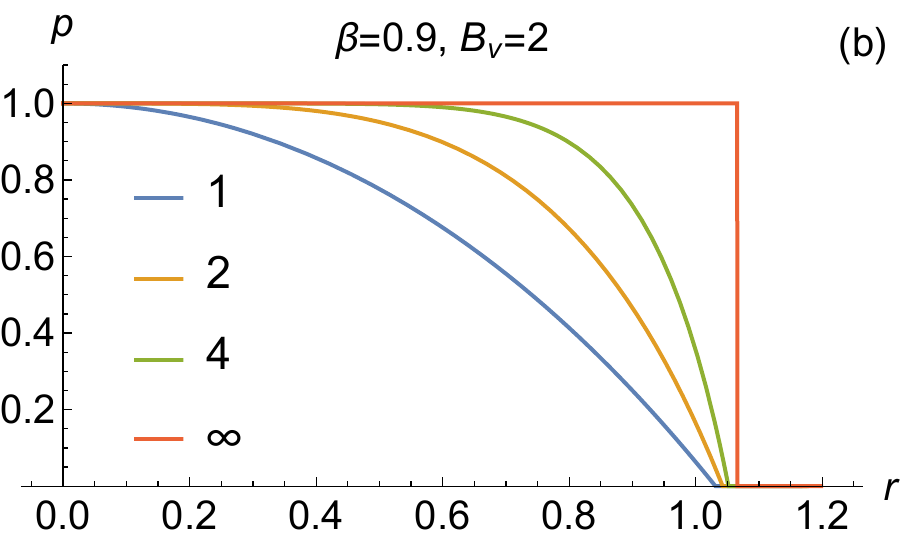}\hfil
  \includegraphics[width=0.3\linewidth]{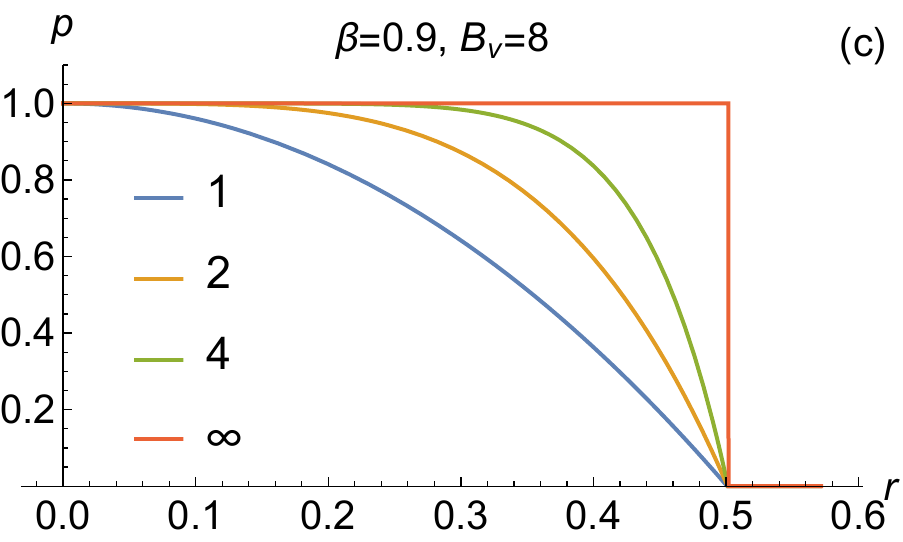}
  \caption{
    %
    Radial pressure profiles with different indices $k$ (indicated in the figures) at $\beta_{}=0.9$ in three plasma sections, where $B_{v}=\{1,2,8\}$.
  }
  \label{fig:p_vs_r}
\end{figure*}

%
The calculations were performed in the Wolfram \emph{Mathematica}$^{\copyright}$ for radial pressure profiles of the form
    \begin{equation}
    \label{4:12}
    f_{k}(\psi) = 1 - \psi^{k}
    \qquad
    (0\leq \psi \leq 1)
    \end{equation}
%
for four indices $k=\{1,2,4,\infty \}$.
Function $f_{1}$ describes the most smooth pressure profile. For $\beta_{}/B_{v}^{2}\ll1$ it approximately gives the parabolic dependence of the pressure $p$ on the radial coordinate $r$. The larger the index $k$, the flatter the radial distribution $p$ near the axis of the plasma column and the steeper it will be near the boundary of the column. Index $k=\infty $ corresponds to the sharp-boundary pressure profile which can be written in terms of a $\theta$-function such that $\theta(x)=0$ for $x<0$ and $\theta(x)=1$ for $x>0$:
    \begin{equation}
    \label{4:13}
    f_{\infty}(\psi) = \theta(1 - \psi)
    .
    \end{equation}
%
Figure \ref{fig:p_vs_r} shows the radial pressure profiles in three sections of the plasma column for the above four values of the index $k$. Wolfram \emph{Mathematica}$^{\copyright}$ was able to calculate the integrals \eqref{4:09} and \eqref{4:11} in analytical form, and the integrals with $k=4$ are expressed in terms of the hypergeometric function ${_2F_1}$:
%
    \begin{subequations}
    \label{4:17}
    \begin{gather}
    \label{4:17-1}
    \frac{a_{1}^{2}}{2}
    =
    \frac{2}{B_{v}+\sqrt{B_{v}^2-\beta_{} }}
    ,\\
    \label{4:17-2}
    \frac{a_{2}^{2}}{2}
    =
    \frac{1}{\sqrt{\beta_{}}}
    \sinh^{-1}
        \left(
            \sqrt{\frac{\beta}{B_{v}^2-\beta}}
        \right)
    ,\\
    \label{4:17-3}
    \frac{a_{4}^{2}}{2}
    =
    \frac{1}{\sqrt{B_{v}^2-\beta_{} }}\,
    {
        {_2F_1}\left(
            \frac{1}{4},\frac{1}{2};
            \frac{5}{4};
            -\frac{\beta_{} }{B_{v}^2-\beta_{}}
        \right)
    }
    ,\\
    \label{4:17-4}
    \frac{a_{\infty }^{2}}{2}
    =
    \frac{1}{\sqrt{B_{v}^2-\beta_{} }}
    ;
    \end{gather}
    \end{subequations}
    \begin{subequations}
    \label{4:18}
    \begin{gather}
    \label{4:18-1}
    \frac{a^{2}_{1}}{2}
    \mean{\overline{p}}_{1}
    =
    -\frac{-2 B_{v}^3+2 B_{v}^2 \sqrt{B_{v}^2-\beta_{} }+\beta_{}  \sqrt{B_{v}^2-\beta_{} }}{3 \beta_{} }
    ,\\
    \label{4:18-2}
    \frac{a^{2}_{2}}{2}
    \mean{\overline{p}}_{2}
    =
    \frac{1}{4} \left(
        \frac{
            \left(\beta_{} +B_{v}^2\right) \coth^{-1}\left(
                {B_{v}}/{\sqrt{\beta_{}}}
            \right)
        }{\sqrt{\beta_{} }}-B_{v}\right)
    ,\\
    \label{4:18-4}
    \frac{a^{2}_{4}}{2}
    \mean{\overline{p}}_{4}
    =
    \frac{1}{6} \left(
        \frac{
            \left(2 \beta_{} +B_{v}^2\right)
        }{
            \sqrt{B_{v}^2-\beta_{} }
        }
            {_2F_1}\left(
                \frac{1}{4},\frac{1}{2};
                \frac{5}{4};
                -\frac{\beta_{} }{B_{v}^2-\beta_{}}
            \right)
        -
        B_{v}\right)
    ,\\
    \label{4:18-inf}
    \frac{a^{2}_{\infty }}{2}
    \mean{\overline{p}}_{\infty }
    =
    \frac{\beta_{} }{2 \sqrt{B_{v}^2-\beta_{} }}
    .
    \end{gather}
    \end{subequations}
Calculating the coefficient
    \begin{equation}
    \label{4:19}
    A=
    \left(
        \frac{B_{v}'}{B_{v}} + \frac{2a'}{a}
    \right)
    ,
    \end{equation}
%
reveals that it is proportional to the derivative of the vacuum magnetic field:
    \begin{subequations}
    \label{4:20}
    \begin{gather}
    \label{4:20-1}
    A_{1}
    =
    \left(\frac{1}{B_{v}}-\frac{1}{\sqrt{B_{v}^2-\beta_{} }}\right) B_{v}'
    ,\\
    \label{4:20-2}
    A_{2}
    =
    \left(
        \frac{1}{B_{v}}
        -
        \frac{\sqrt{\beta_{} }}{
            \left(B_{v}^2-\beta_{}\right)
            \text{csch}^{-1}
            \left(
                \sqrt{{B_{v}^2}/{\beta_{} }-1}
            \right)
        }
    \right)
    B_{v}'
    ,\\
    \label{4:20-4}
    A_{4}
    =
    \frac{1}{2}
    \left(
        \frac{B_{v}^2-2\beta_{}}{B_{v}^3-\beta_{}B_{v}}
        -
        \frac{1}{
            \sqrt{B_{v}^2-\beta_{} }
            \,
            {_2F_1}\left(
                \frac{1}{4},\frac{1}{2};
                \frac{5}{4};
                -\frac{\beta_{} }{B_{v}^2-\beta_{}}
            \right)
        }
    \right)
    B_{v}'
    ,\\
    \label{4:20-inf}
    A_{\infty }
    =
    \frac{\beta_{}  B_{v}'}{\beta_{}  B_{v}-B_{v}^3}
    .
    \end{gather}
    \end{subequations}

%
The authors of the publications cited above used various axial profiles of the vacuum magnetic field $B_{v}$ in their calculations. For example,
\emph{D'Ippolito and Myra} \cite{DIppolitoMyra1984PF_27_2256} studied plasma stabilization by some external force of unspecified nature by modeling the magnetic field in a tandem trap with an interpolation function that approximately replicated the real vacuum field. \emph{Kesner} \cite{Kesner1985NF_25_275} studied wall stabilization in the $\Lambda\to\infty $ limit by simulating the magnetic field with a parabola. \emph{Li, Kesner and Lane} \cite{LiKesnerLane1985NF_25_907, LiKesnerLane1987NF_27_101} modeled the magnetic field as a sum of a constant and a cosine. \emph{Li, Kesner and LoDestro} \cite{LiKesnerLoDestro1987NF_27_1259} did about the same. Binding to only one specific field model in these works does not allow one to find out how the axial profile of the magnetic field should be modified in order to lower the critical beta and thereby simplify the transition to a stable plasma confinement regime.

\begin{figure*}
  \centering
  \includegraphics[width=0.3\linewidth]{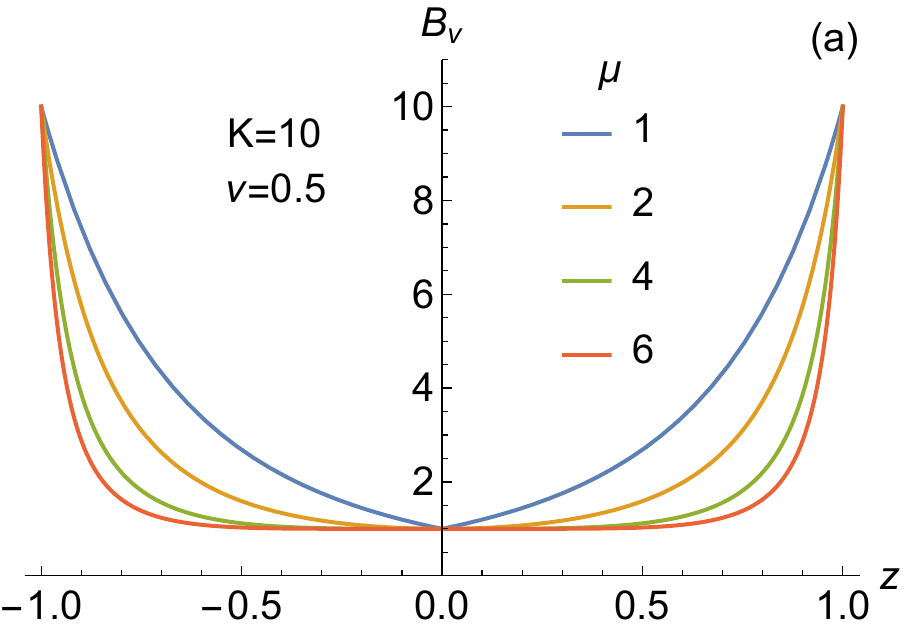}\hfil
  \includegraphics[width=0.3\linewidth]{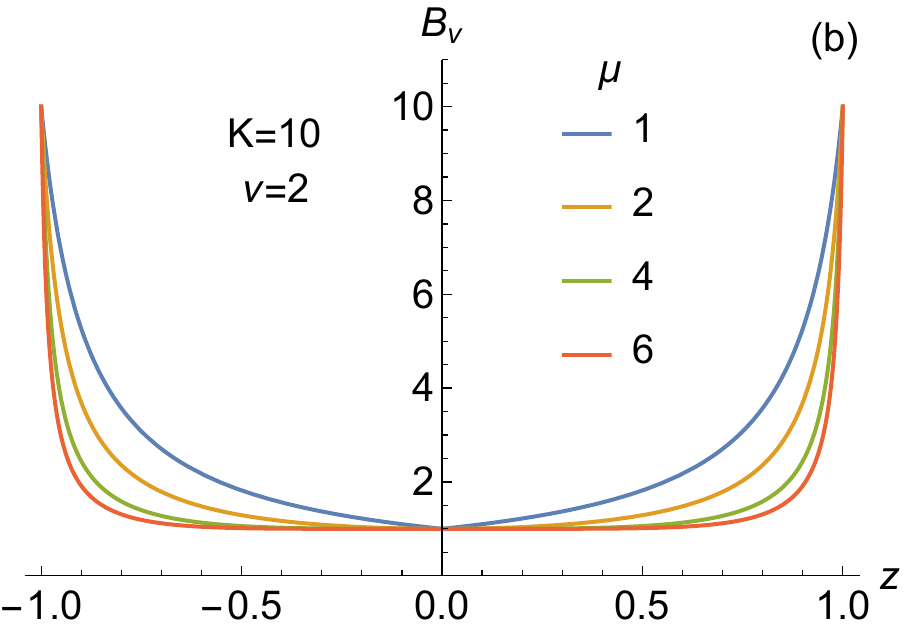}\hfil
  \includegraphics[width=0.3\linewidth]{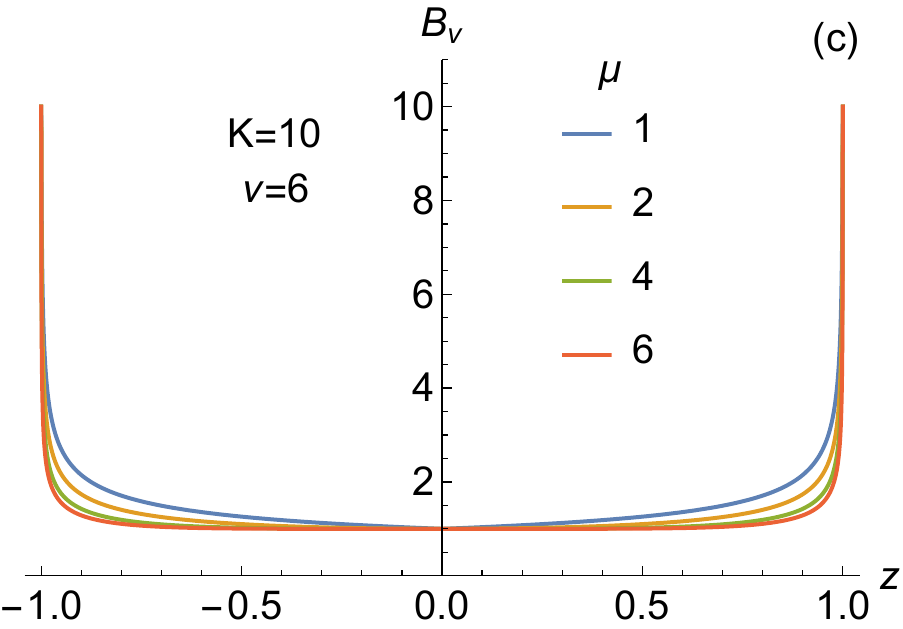}
  \caption{
    %
    Axial profiles of the magnetic field corresponding to different values of the parameters $\mu$ and $\nu$ (shown in the figures) for the same mirror ratio $K=10$.
  }
  \label{fig:GM-Bv}
\end{figure*}

%
To investigate the dependence of critical beta on the axial profile of the magnetic field, we used two models. In the first model, the vacuum magnetic field was given by a three-parameter family of functions
    \begin{equation}
    \label{4:31}
    B_{v}(z)
    =
    \left[
        1
        -
        \left(
            1 - K^{-\nu/2}
        \right)
        |z|^{\mu}
    \right]^{-2/\nu }
    ,
    \end{equation}
%
which depended on the mirror ratio $K$ and two indices $\mu$ and $\nu$. Previously, such a family was used by Mirnov and Bushkova \cite{BushkovaMirnov1986VANT_2_19e}, as well as by ourselves in a recent paper \cite{Kotelnikov+2021PST_24_015102}. The meaning of the parameter $K=B_{v}(\pm1)/B_{v}(0)$ is obvious from its name, and the indices $\mu$ and $\nu$ determine the width and “steepness” of magnetic mirrors. The calculations reported below were performed for the following combinations of parameters: $K=\{20,16,12,8,4\}$, $\mu=\{1,2,4,6\}$, $\nu=\{0.5 ,2,6\}$. Figure \ref{fig:GM-Bv} shows the profiles of the vacuum magnetic field for different combinations of the $\mu$ and $\nu$ parameters. It is easy to see that both with an increase in $\mu$ and with an increase in $\nu$, the profile steepens near the magnetic mirrors while the quasi-homogeneous region at the center of the trap expands.

The family of functions \eqref{4:31} has the peculiarity that the derivative $B_{v}'$ does not vanish at the ends of the integration interval $z=\pm1$, as it was assumed when deriving Eq.~\eqref{4:11}. Physically, this should mean that magnetic coil of such a small size is installed in the mirror throat that, on the scale under consideration, the coil can be considered "point". The method for correctly taking into account this and one more feature of the model \eqref{4:31} is described in Appendix \ref{A1}. It is also useful to point out here that the combination $\mu=1$, $\nu=2$ minimizes the absolute value of the integral in the Rosenbluth-Longmire criterion \cite{RosenbluthLongmire1957AnnPhys_1_120}, which determines the stability condition for flute oscillations in open traps (see \cite{ Kotelnikov2021V2e}).


\begin{table}
  \centering
\begin{equation*}
\begin{array}{|c|c|cccc|}
\hline
 \text{k} & \nu \backslash \backslash \mu  & 1 & 2 & 4 & 6 \\
\hline
 \text{} & 0.5 & \num{0.969577} & \text{N/F} & \text{N/F} & \text{N/F} \\
 1 & 2 & \num{0.989506} & \text{N/F} & \text{N/F} & \text{N/F} \\
\text{} & 6 & \text{N/F} & \text{N/F} & \text{N/F} & \text{N/F} \\
\hline
 \text{} & 0.5 & \num{0.886506} & \num{0.976543} & \num{0.99766} & \num{0.999706} \\
 2 & 2 & \num{0.929788} & \num{0.993294} & \num{0.999931} & 1. \\
 \text{} & 6 & \num{0.998716} & 1. & \text{N/F} & \text{N/F} \\
\hline
 \text{} & 0.5 & \num{0.831537} & \num{0.941871} & \num{0.980161} & \num{0.988949} \\
 4 & 2 & \num{0.886061} & \num{0.974189} & \num{0.995123} & \num{0.998254} \\
 \text{} & 6 & \num{0.995489} & \num{0.999953} & 1. & \text{N/F} \\
\hline
 \text{} & 0.5 & \bnum{0.767052} & \num{0.894361} & \num{0.946909} & \num{0.961585} \\
 \infty  & 2 & \num{0.833547} & \num{0.943741} & \num{0.979146} & \num{0.987335} \\
 \text{} & 6 & \num{0.990581} & \num{0.999692} & \num{0.999997} & 1. \\
\hline
\end{array}
\end{equation*}
  \caption{
    $\beta_{\text{crit}}$ for an isotropic plasma in a magnetic field \eqref{4:31} at
     $K=20$ and $\Lambda =\infty$.
     Minimum value
         $\beta_{\min}=\num{0.767052}$
     is achieved for $k=\infty $, $\mu=1$, $\nu=0.5$.
  }\label{tbl:K=20}
  \centering
\begin{equation*}
\begin{array}{|c|c|cccc|}
\hline
 \text{k} & \nu \backslash \backslash \mu  & 1 & 2 & 4 & 6 \\
\hline
 \text{} & 0.5 & \num{0.969426} & \text{N/F} & \text{N/F} & \text{N/F} \\
 1 & 2 & \num{0.989124} & \text{N/F} & \text{N/F} & \text{N/F} \\
 \text{} & 6 & \text{N/F} & \text{N/F} & \text{N/F} & \text{N/F} \\
\hline
 \text{} & 0.5 & \num{0.886211} & \num{0.976236} & \num{0.997553} & \num{0.999677} \\
 2 & 2 & \num{0.928691} & \num{0.99289} & \num{0.999914} & 1. \\
 \text{} & 6 & \num{0.998201} & \num{0.999999} & \text{N/F} & \text{N/F} \\
\hline
 \text{} & 0.5 & \num{0.831166} & \num{0.941338} & \num{0.979741} & \num{0.988616} \\
 4 & 2 & \num{0.88457} & \num{0.973184} & \num{0.99471} & \num{0.998038} \\
 \text{} & 6 & \num{0.994062} & \num{0.999919} & 1. & \text{N/F} \\
\hline
 \text{} & 0.5 & \bnum{0.766592} & \num{0.893567} & \num{0.9461} & \num{0.960821} \\
 \infty  & 2 & \num{0.831589} & \num{0.941978} & \num{0.977951} & \num{0.986398} \\
 \text{} & 6 & \num{0.987895} & \num{0.999495} & \num{0.999993} & 1. \\
\hline
\end{array}
\end{equation*}
  \caption{
    Critical beta for an isotropic plasma in a magnetic field \eqref{4:31} at
     $K=16$ and $\Lambda =\infty$.
     Minimum value
     $\beta_{\min}=\num{0.766592}$
     is achieved for $k=\infty $, $\mu=1$, $\nu=0.5$.
  }\label{tbl:K=16}
  \centering
\begin{equation*}
\begin{array}{|c|c|cccc|}
\hline
 \text{k} & \nu \backslash \backslash \mu  & 1 & 2 & 4 & 6 \\
\hline
 \text{} & 0.5 & \num{0.969118} & \text{N/F} & \text{N/F} & \text{N/F} \\
 1 & 2 & \num{0.988454} & \text{N/F} & \text{N/F} & \text{N/F} \\
 \text{} & 6 & \text{N/F} & \text{N/F} & \text{N/F} & \text{N/F} \\
\hline
 \text{} & 0.5 & \num{0.885611} & \num{0.975622} & \num{0.997334} & \num{0.999615} \\
 2 & 2 & \num{0.926798} & \num{0.992157} & \num{0.999879} & \num{0.999999} \\
 \text{} & 6 & \num{0.997185} & \num{0.999997} & \text{N/F} & \text{N/F} \\
\hline
 \text{} & 0.5 & \num{0.83041} & \num{0.940276} & \num{0.978901} & \num{0.987946} \\
 4 & 2 & \num{0.882007} & \num{0.971405} & \num{0.993941} & \num{0.997619} \\
 \text{} & 6 & \num{0.991444} & \num{0.999829} & 1. & \text{N/F} \\
\hline
 \text{} & 0.5 & \bnum{0.765656} & \num{0.891987} & \num{0.944498} & \num{0.959308} \\
 \infty  & 2 & \num{0.828229} & \num{0.938884} & \num{0.975797} & \num{0.98468} \\
 \text{} & 6 & \num{0.983111} & \num{0.99903} & \num{0.999977} & \num{0.999998} \\
\hline
\end{array}
\end{equation*}
  \caption{
    Critical beta for an isotropic plasma in a magnetic field \eqref{4:31} at
     $K=12$ and $\Lambda =\infty$.
     Minimum value
     $\beta_{\min}=\num{0.765656}$
     is achieved for $k=\infty $, $\mu=1$, $\nu=0.5$.
  }\label{tbl:K=12}
\end{table}

\begin{table}
  \centering
\begin{equation*}
\begin{array}{|c|c|cccc|}
\hline
 \text{k} & \nu \backslash \backslash \mu  & 1 & 2 & 4 & 6 \\
\hline
 \text{} & 0.5 & \num{0.968295} & \text{N/F} & \text{N/F} & \text{N/F} \\
 1 & 2 & \num{0.98697} & \text{N/F} & \text{N/F} & \text{N/F} \\
 \text{} & 6 & 1. & \text{N/F} & \text{N/F} & \text{N/F} \\
 \hline
\text{} & 0.5 & \num{0.884018} & \num{0.974017} & \num{0.996725} & \num{0.999421} \\
 2 & 2 & \num{0.922754} & \num{0.990435} & \num{0.999758} & \num{0.999996} \\
 \text{} & 6 & \num{0.994532} & \num{0.999986} & \text{N/F} & \text{N/F} \\
 \hline
\text{} & 0.5 & \num{0.82841} & \num{0.937536} & \num{0.976714} & \num{0.986174} \\
 4 & 2 & \num{0.876563} & \num{0.967413} & \num{0.992044} & \num{0.996504} \\
 \text{} & 6 & \num{0.9853} & \num{0.999478} & \num{0.999996} & 1. \\
\hline
 \text{} & 0.5 & \bnum{0.763185} & \num{0.887935} & \num{0.940394} & \num{0.955414} \\
 \infty  & 2 & \num{0.821129} & \num{0.932068} & \num{0.970801} & \num{0.98057} \\
 \text{} & 6 & \num{0.97234} & \num{0.997444} & \num{0.999857} & \num{0.999979} \\
\hline
\end{array}
\end{equation*}
  \caption{
    Critical beta for an isotropic plasma in a magnetic field \eqref{4:31} at
     $K=8$ and $\Lambda =\infty$.
     Minimum value
     $\beta_{\min}=\num{0.763185}$
     is achieved for $k=\infty $, $\mu=1$, $\nu=0.5$.
  }\label{tbl:K=8}
  \centering
\begin{equation*}
\begin{array}{|c|c|cccc|}
\hline
 \text{k} & \nu \backslash \backslash \mu  & 1 & 2 & 4 & 6 \\
\hline
 \text{} & 0.5 & \num{0.964113} & \text{N/F} & \text{N/F} & \text{N/F} \\
 1 & 2 & \num{0.981067} & \text{N/F} & \text{N/F} & \text{N/F} \\
 \text{} & 6 & \num{0.999894} & \text{N/F} & \text{N/F} & \text{N/F} \\
\hline
 \text{} & 0.5 & \num{0.876139} & \num{0.965939} & \num{0.992846} & \num{0.997602} \\
 2 & 2 & \num{0.908158} & \num{0.982414} & \num{0.99832} & \num{0.999791} \\
 \text{} & 6 & \num{0.980788} & \num{0.9996} & 1. & \text{N/F} \\
\hline
 \text{} & 0.5 & \num{0.818595} & \num{0.924308} & \num{0.965448} & \num{0.976498} \\
 4 & 2 & \num{0.857282} & \num{0.950967} & \num{0.982075} & \num{0.989404} \\
 \text{} & 6 & \num{0.95888} & \num{0.995161} & \num{0.999671} & \num{0.999949} \\
\hline
 \text{} & 0.5 & \bnum{0.751161} & \num{0.868862} & \num{0.920574} & \num{0.936167} \\
 \infty  & 2 & \num{0.796373} & \num{0.905503} & \num{0.948568} & \num{0.960802} \\
 \text{} & 6 & \num{0.929436} & \num{0.983598} & \num{0.996047} & \num{0.998187} \\
\hline
\end{array}
\end{equation*}
  \caption{
    Critical beta for an isotropic plasma in a magnetic field \eqref{4:31} at
     $K=4$ and $\Lambda =\infty$.
     Minimum value
     $\beta_{\min}=\num{0.751161}$
     is achieved for $k=\infty $, $\mu=1$, $\nu=0.5$.
  }\label{tbl:K=4}
  \centering
\begin{equation*}
\begin{array}{|c|c|cccc|}
\hline
 \text{k} & \nu \backslash \backslash \mu  & 1 & 2 & 4 & 6 \\
\hline
 \text{} & 0.5 & \num{0.944517} & \num{0.997926} & \text{N/F} & \text{N/F} \\
 1 & 2 & \num{0.957531} & \num{0.999963} & \text{N/F} & \text{N/F} \\
 \text{} & 6 & \num{0.987054} & \text{N/F} & \text{N/F} & \text{N/F} \\
\hline
 \text{} & 0.5 & \num{0.842901} & \num{0.925434} & \num{0.959601} & \num{0.96962} \\
 2 & 2 & \num{0.862057} & \num{0.939736} & \num{0.970102} & \num{0.978678} \\
 \text{} & 6 & \num{0.916647} & \num{0.974858} & \num{0.992704} & \num{0.996555} \\
\hline
 \text{} & 0.5 & \num{0.778691} & \num{0.866595} & \num{0.905872} & \num{0.918198} \\
 4 & 2 & \num{0.799898} & \num{0.884208} & \num{0.920471} & \num{0.931682} \\
 \text{} & 6 & \num{0.863613} & \num{0.932931} & \num{0.959835} & \num{0.967726} \\
\hline
 \text{} & 0.5 & \bnum{0.704229} & \num{0.793394} & \num{0.834532} & \num{0.847731} \\
 \infty  & 2 & \num{0.726839} & \num{0.813216} & \num{0.851734} & \num{0.863953} \\
 \text{} & 6 & \num{0.797425} & \num{0.871675} & \num{0.902472} & \num{0.912002} \\
\hline
\end{array}
\end{equation*}
  \caption{
    Critical beta for an isotropic plasma in a magnetic field \eqref{4:31} at
     $K=2$ and $\Lambda =\infty$.
     Minimum value
     $\beta_{\min}=\num{0.704229}$
     is achieved for $k=\infty $, $\mu=1$, $\nu=0.5$.
  }\label{tbl:K=2}
\end{table}

The search for the roots of Eq.~\eqref{4:10} was carried out using the \texttt{FindRoot} utility built into the Wolfram \emph{Mathematica}$^{\copyright}$ language. \texttt{FindRoot} searches for the first root near the initial guess $\beta_{\text{start}}$ passed to it. Success or failure in finding the root with this utility depends very much on luck in choosing $\beta_{\text{start}}$. Therefore, as an additional means of searching for roots, the \texttt{RootSearch} package was included, which was developed by Ted Ersek \cite{Ersek_RootSearch}. This package contains a utility of the same name that searches for all roots within a given interval.
%
%
In the final version of our code, the root found by the \texttt{RootSearch} utility was passed as $\beta_{\text{start}}$ to the \texttt{FindRoot} utility to recheck the result of the $\beta_{\text{crit}}$ calculation . In rare cases, when only one of the two utilities found a solution to Eq.~\eqref{4:10}, the code was analyzed in order to improve it. In cases where both \texttt{FindRoot} and \texttt{RootSearch} did not find a solution to Eq.~\eqref{4:10}, it was considered that the solution did not exist.


\begin{figure}
  \centering
  \includegraphics[width=\linewidth]{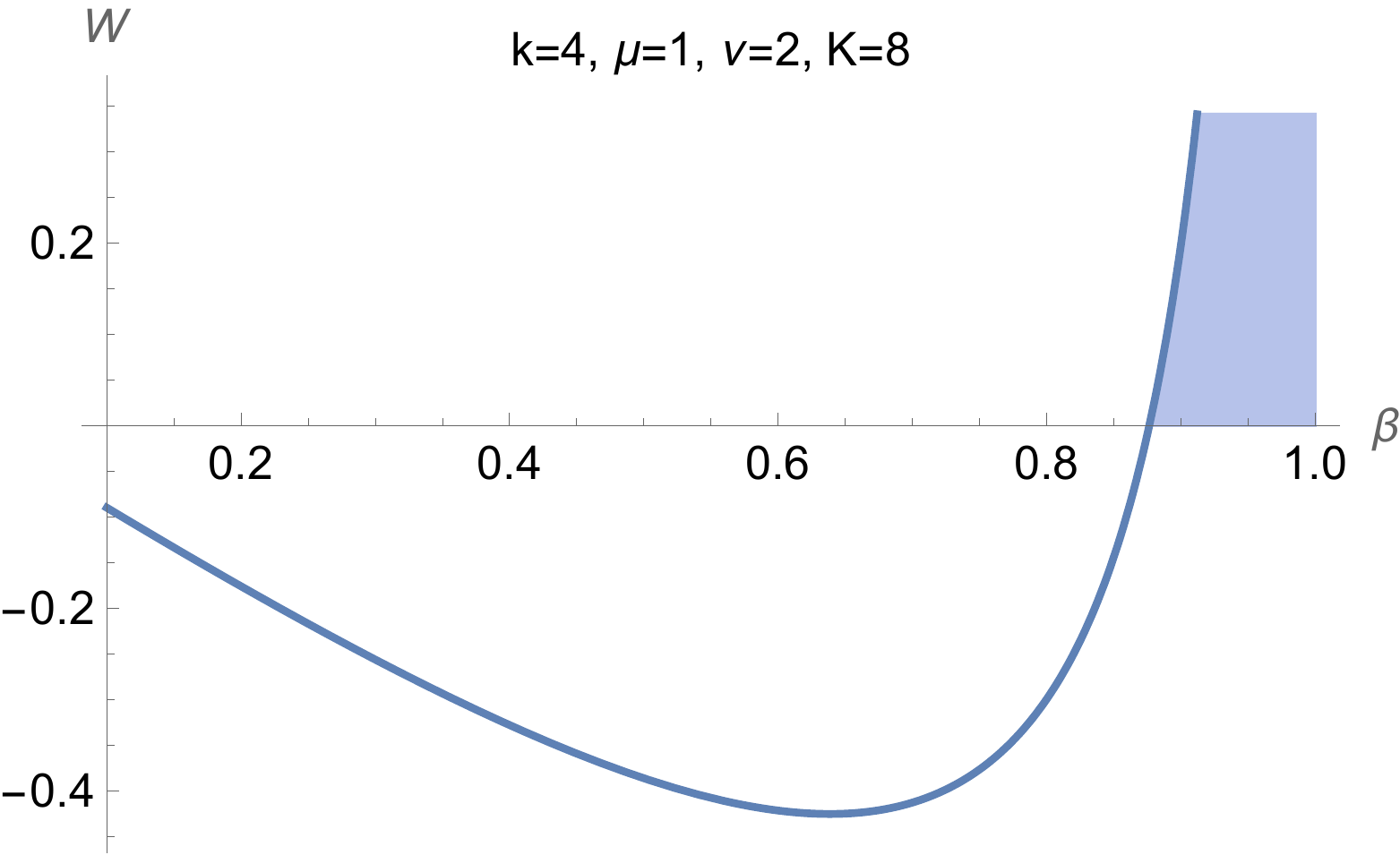}
  \caption{
    %
    An example of the dependence of the integral \eqref{4:11} on $\beta_{}$ for a field of the form \eqref{4:31}. The region of stability $W>0$ for $\beta_{} > \beta_{\text{crit}}$ is shaded.
  }
  \label{fig:W1_k4_q2_K8}
\end{figure}

%
The results of the numerical solution of Eq.~\eqref{4:10} are collected in tables \ref{tbl:K=20}–\ref{tbl:K=2}. First of all, it is useful to check that the calculated values of $\beta_{\text{crit}}$ do indeed indicate the lower margin of the stability zone. To do this, it suffices to study the dependence of the integral \eqref{4:11} on $\beta_{}$. An example of such a dependence is shown in Figure~\ref{fig:W1_k4_q2_K8}. It proves that there is only one stability zone $W>0$ and that it is located in the region $\beta_{}>\beta_{\text{crit}}$, where $\beta_{\text{crit}}$ is the root of Eq.~\eqref{4:10}.

%
Each table is made for one fixed value of the mirror ratio $K$. Within each individual table, it is not difficult to detect a trend towards a decrease in the critical beta with an increase in the steepness of the radial pressure profile as the index $k$ increases from $k=1$ to $k=\infty$ for a fixed pair of indices $\mu$ (running horizontally from $1$ to $6$) and $\nu$ (vertically down from $0.5$ to $6$). The abbreviation N/F instead of a number says that the root was not found. This can mean both that the root does not exist, or that it exists but is less than $1$ by less than $10^{-6}$. From a practical point of view, it's all the same: it's hard to imagine that in a real experiment one can get so close to the theoretical limit $\beta_{}=1$.

%
Further, we see that the critical beta increases both as the index $\mu$ increases and as the index $\nu$ increases. In other words, stabilization of the rigid ballooning mode is more problematic in traps with short and steep magnetic mirrors. The smallest value of critical beta is reached at $k=\infty $, $\mu=1$ and $\nu=0.5$. It changes slightly within some narrow interval
    from $\beta_{\min}=\num{0.767052}$ for $K=20$
    down to $\beta_{\min}=\num{0.704229}$ for $K=2$.
Comparison of the critical beta values in different tables with the same pairs of indices $\mu$ and $\nu$ also shows that the value of the mirror ratio $K$ has very little effect on the result of calculations in the interval of sufficiently large values of $K$, but begins to decrease more noticeably for $K<4$. However, this fact can hardly be of practical importance, since it is difficult to imagine how an isotropic plasma can be confined in a trap with a small mirror ratio. Some reduction in the set of combinations of indices $k$, $\mu$, $\nu$, for which no solution has been found, with a decrease in $K$, in general, is also only of academic interest.

\begin{table}
  \centering
\begin{equation*}
\begin{array}{|c|c|cccc|}
\hline
 \text{k} & q\backslash \backslash K & 16 & 8 & 4 & 2 \\
\hline
 \text{} & 2 & \num{0.99912} & \num{0.998425} & \num{0.995482} & \num{0.974971} \\
 1 & 4 & \text{N/F} & \text{N/F} & \text{N/F} & \num{0.993741} \\
 \text{} & 8 & \text{N/F} & \text{N/F} & \text{N/F} & \num{0.999212} \\
\hline
 \text{} & 2 & \num{0.936669} & \num{0.93251} & \num{0.920411} & \num{0.874111} \\
 2 & 4 & \num{0.975089} & \num{0.96966} & \num{0.955448} & \num{0.901739} \\
 \text{} & 8 & \num{0.987452} & \num{0.982331} & \num{0.96852} & \num{0.913126} \\
\hline
 \text{} & 2 & \num{0.884435} & \num{0.878896} & \num{0.863139} & \num{0.807071} \\
 4 & 4 & \num{0.933456} & \num{0.924951} & \num{0.90409} & \num{0.835497} \\
 \text{} & 8 & \num{0.952088} & \num{0.94279} & \num{0.920539} & \num{0.847408} \\
\hline
 \text{} & 2 & \num{0.819595} & \num{0.812656} & \num{0.793243} & \bnum{0.728335} \\
 \infty  & 4 & \num{0.876482} & \num{0.865035} & \num{0.837831} & \num{0.756216} \\
 \text{} & 8 & \num{0.899629} & \num{0.886476} & \num{0.856387} & \num{0.767997} \\
\hline
\end{array}
\end{equation*}
  \caption{
  Critical beta for an isotropic plasma in a magnetic field \eqref{4:35} at $\Lambda =\infty$.
     Minimum value
     $\beta_{\min}=0.728335$
     is achieved for $k=\infty $, $q=2$, $K=2$.
  }\label{tbl:Kesner}
\end{table}

%
To be able to compare our calculations with the results of other authors, we also calculated $\beta_{\text{crit}}$ for the second magnetic field model, which is given by a two-parameter family of functions
    \begin{equation}
    \label{4:35}
    B_{v}(z) = 1 + (K-1)\sin^{q}(\pi z/2)
    \end{equation}
%
%
with three index values $q=\{2,4,8\}$. The variant $q=2$ occured in several works, in particular, it was used by \emph{Li, Kesner and Lane}
\cite{LiKesnerLane1985NF_25_907}. The results are presented in table~\ref{tbl:Kesner}. The minimum value is reached at $q=2$. It weakly depends on the mirror ratio in the interval from $K=20$ to $K=4$, but decreases more noticeably with a further decrease in $K$. In particular, $\beta_{\min}=\num{0.728335}$ for $K=2$.
As far as can be judged from the graph in Figure 4 in the article \cite{LiKesnerLane1985NF_25_907}, the critical beta calculated in that article for the stepped plasma profile approximately coincides with the value $\beta_{\text{crit}}=\num{0.812656}$ indicated in our table \ref{tbl:Kesner}  for $k=\infty $, $q=2$, $K=8$.

\begin{figure*}
  \centering
  \includegraphics[width=0.3\linewidth]{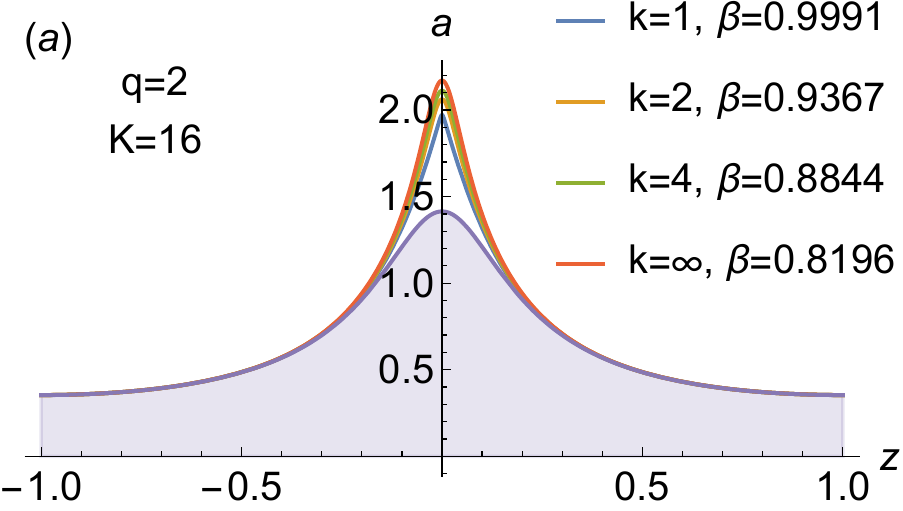}\hfil
  \includegraphics[width=0.3\linewidth]{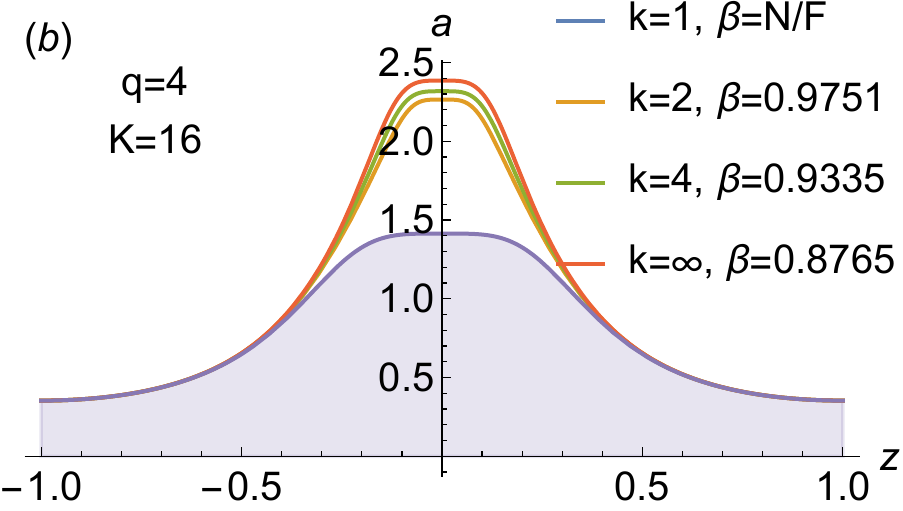}\hfil
  \includegraphics[width=0.3\linewidth]{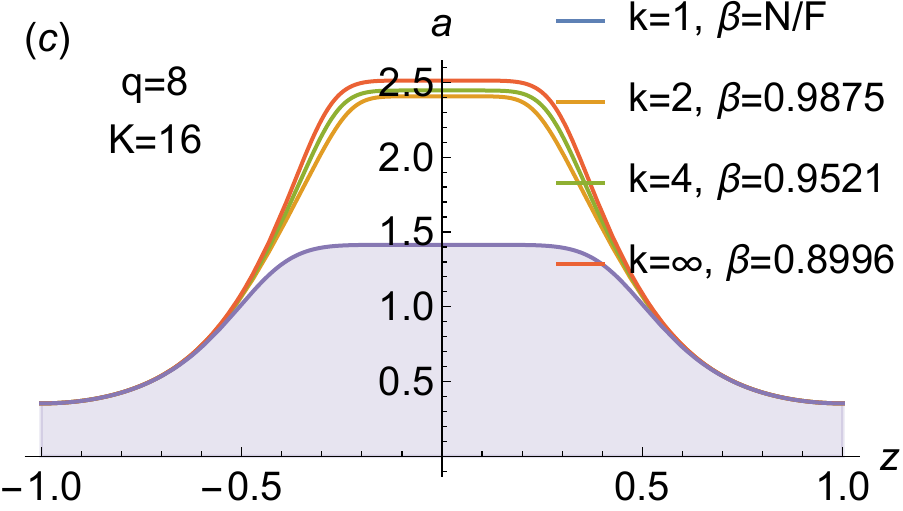}
  \caption{
    %
    The profile of the plasma boundary in the field in the form \eqref{4:35} for different values of the parameter $q$ (shown in the figures) and critical values of beta for different pressure profiles with different parameters $k$ (shown in the figures). The area occupied by plasma at $\beta =0$ is shaded.
  }\label{fig:GM-a_vs_z}
\end{figure*}

%
Unlike the first field model \eqref{4:31}, functions \eqref{4:35} are smooth everywhere and have no kinks. But even in the absence of such a kink on the vacuum field profile $B_{v}(z)$, on the profile of the plasma boundary $a(z)$ near the median plane $z=0$, a ``swell'' is formed in the form of a ``thorn'' with a large curvature on spearhead. An example of such a ``spike'' for $q=2$ is shown in Figure~\ref{fig:GM-a_vs_z}(a). At $q=8$ the ``thorn'' expands, forming a diamagnetic ``bubble'' named after Beklemishev \cite{Beklemishev2016PoP_23_082506}, as in Figure~\ref{fig:GM-a_vs_z}(c). The plots of the plasma boundary $a(z)$ in Figure~\ref{fig:GM-a_vs_z} are plotted for different values of $k$ and corresponding to them values of $\beta_{\text{crit}}$, but with the same magnetic flux $\psi=\psi_{a}=1$ captured in plasma. Interestingly, such plots $a(z)$ almost coincide, although the values of $\beta_{\text{crit}}$ for different $k$ differ quite significantly.

\begin{figure*}
  \centering
  \includegraphics[width=0.3\linewidth]{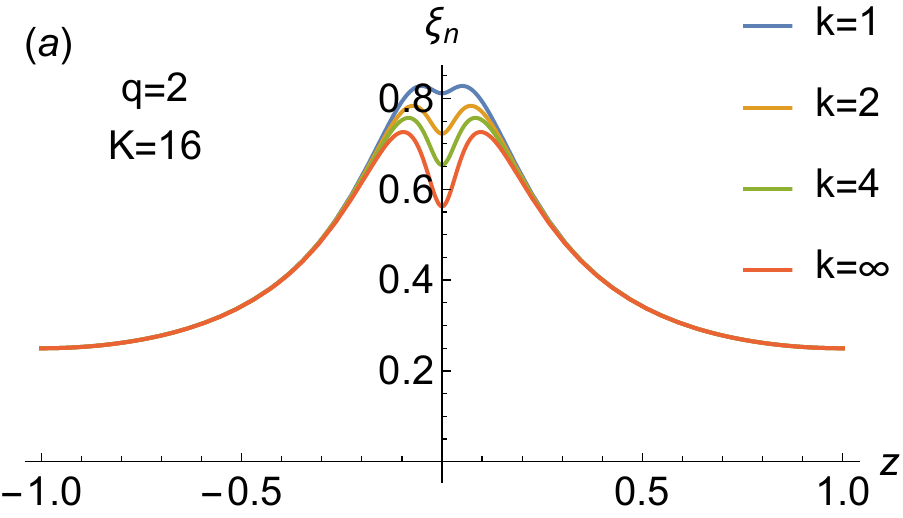}\hfil
  \includegraphics[width=0.3\linewidth]{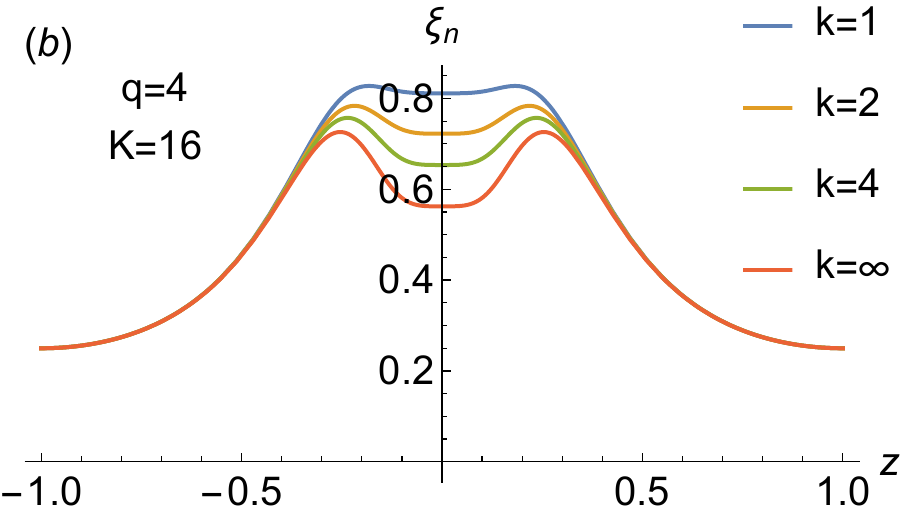}\hfil
  \includegraphics[width=0.3\linewidth]{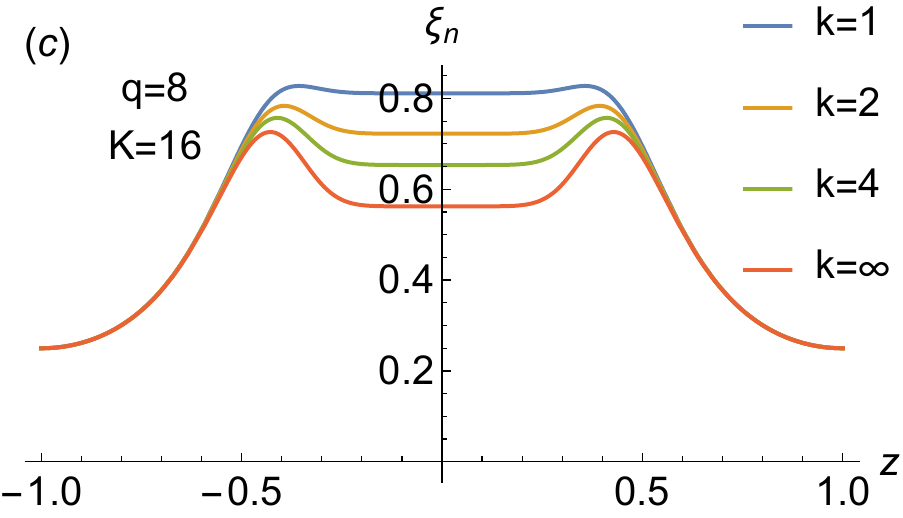}
  \caption{
    %
    Displacement profile $\xi_{n}(z)$ of the plasma column in the field \eqref{4:35} for $\beta =0.9$, $\Lambda \to \infty $ and various values of the parameter $q$ and $k $ (indicated in the figures).
  }
  \label{fig:GM-ksi_vs_z}
\end{figure*}
%
The displacement profile of the plasma column $\xi_{n}(z)$ is shown in Figure~\ref{fig:GM-ksi_vs_z} for the same value $\beta=0.9$ for all radial pressure profiles $k$. We emphasize that the displacement is not constant, although, as mentioned above, $\phi(z)=\const$ for $\Lambda \to\infty$. At the critical values of beta indicated in table \ref{tbl:Kesner} and in figure \ref{fig:GM-a_vs_z}, the displacement profiles would be practically the same for all $k$, since the profiles of the plasma boundary $a(z) $ in Figure~\ref{fig:GM-a_vs_z} practically coincide.

%
The next section describes the method and results of solving the LoDestro equation \eqref{3:01} with a finite value of the parameter $\Lambda $. We used the tables \ref{tbl:K=20}–\ref{tbl:K=2} and \ref{tbl:Kesner} to check convergence of the method we used for large values of the parameter $\Lambda $.

\section{Case of finite vacuum gap}\label{s5}

Taking into account the symmetry of the magnetic field with respect to the median plane $z=0$, it suffices to find a solution to the equation \eqref{3:01} at half the distance between the magnetic mirrors, for example, in the interval $0 \leq z \leq 1$. As we explained briefly in section \ref{s4} and in detail in Appendix \ref{A1}, when implementing the first model of the magnetic field \eqref{4:31}, the coefficients of the equation can be singular  at the end points of this interval due to the presence of delta functions $\delta(z)$ and $\delta(z-1)$. Such singularities can be smoothed out using one of the methods proposed in Appendix \ref{A1}, or the boundary conditions can be moved inside the interval to the points $z=0+$ (slightly to the right of $z=0$) and $z=1-$ (slightly to the left $z=1$), as is done in Appendix \ref{A2}. Further in this and the next sections, we formulate the boundary conditions at the end points of the interval $0\leq z \leq 1$, omitting the details of the transfer of the boundary conditions inside the interval $0+ < z < 1-$, where the coefficients of the equation \eqref{3:01} do not have peculiarities.

We used the built-in utility \texttt{Parametric\-NDSolve\-Value} to find a marginal solution to the ordinary differential equation \eqref{3:01} for $\omega=0$ in the Wolfram \emph{Mathematica}$^{\copyright}$ system. It returns a reference $pf$ to the interpolation function of the $z$ coordinate, which also depends on the free parameters $\beta_{}$ and $\Lambda$. The other parameters ($K$, $\mu$ and $\nu$ or $K$ and $q$) were given as numbers.

%
The boundary conditions for the \texttt{Parametric\-NDSolve\-Value} utility were specified on the left boundary as
    \begin{equation}
    \label{5:01}
    \phi(0)=1,
    \qquad
    \phi'(0)=0
    \end{equation}
%
It would be a mistake to specify a pair of boundary conditions $\phi'(0)=0$ and $\phi'(1)=0$, because then the \texttt{Parametric\-NDSolve\-Value} utility would only find a trivial solution $ \phi(z)\equiv0$. Therefore, the normalization condition $\phi(0)=1$ or $\phi(1)=1$ is absolutely necessary. However, it would also be a mistake to specify the third boundary condition $\phi'(1)=0$ on the right boundary in addition to \eqref{5:01}, since a second-order ordinary differential equation with three boundary conditions does not have a solution except for some ``eigenvalue'' of $\beta $ at given value of $\Lambda $. In the terminology used above, this eigenvalue is the critical beta $\beta_{\text{crit}}$.

%
To calculate $\beta_{\text{crit}}$ for a given value of the $\Lambda$ parameter, we essentially used the shooting method. In the classical implementation, this method consists in the fact that the differential equation is numerically integrated for a certain numerical value $\beta_{}$ and given boundary conditions on one boundary of the interval. The found solution is checked on the opposite boundary, comparing it with the ``target'', that is, with the boundary condition at this boundary. In the next step, the given value of $\beta_{}$ is adjusted with the intent to ``hit the target''. This is where the name of the ``shooting'' method comes from.
In Wolfram \emph{Mathematica}$^{\copyright}$ implementation of the shooting method, there is no need to repeatedly integrate the differential equation with each new value of $\beta_{}$, since the \texttt{Parametric\-NDSolve\-Value} utility has already done everything.

%
As mentioned above, this utility returns a reference $pf$ to the interpolation function, which is the solution of the equation passed to it with the above boundary conditions \eqref{5:01}. In the Wolfram \emph{Mathematica}$^{\copyright}$, function arguments are written in square brackets, so $pf[\beta,\Lambda][z]$ denotes the solution at $z$ for specific numerical values of the parameters $\beta$ and $\Lambda$. Accordingly, the analogue of the derivative $\phi'(z)$ is written as $pf[\beta,\Lambda]'[z]$. So to find the critical beta, it is enough to pass the equation
    \begin{equation}
   \label{5:02}
    pf[\beta,\Lambda]'[1]=0
    \end{equation}
to the \texttt{FindRoot} or \texttt{RootSearch} utility that we mentioned in the previous section. In fact, we used both of these utilities, calculating the root of Eq.~\eqref{5:02} twice. We concluded that the root did not exist only if both utilities did not find a solution to Eq.~\eqref{5:02}.

%
%
%

%
Calculations were made for both models of the magnetic field and those combinations of parameters $k$, $K$, $\mu=$, $\nu$, which are listed in tables \ref{tbl:K=20}–\ref{tbl:Kesner}. The $\Lambda$ parameter could take discrete values $\Lambda=\{1,1.01,1.02,1.05, 1.1,\ldots 500,1000\}$ in the range from $\Lambda =1$ to $\Lambda=1000$. For $\Lambda=500$, the critical beta value we calculated differed from the value found in the previous section for $\Lambda=\infty$ only in the fifth decimal place.


%

\begin{figure}
  \centering
%
  \includegraphics[width=0.49\linewidth]{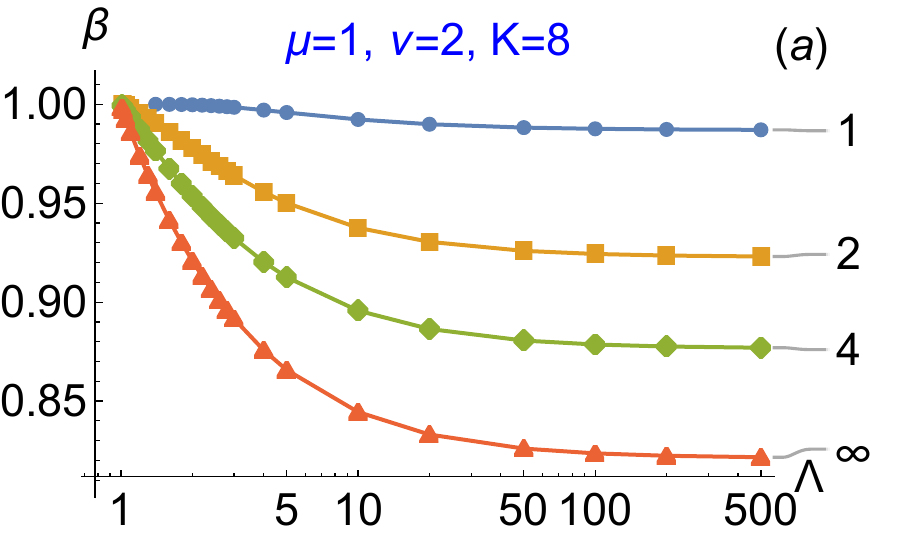}
  \hfill
  \includegraphics[width=0.49\linewidth]{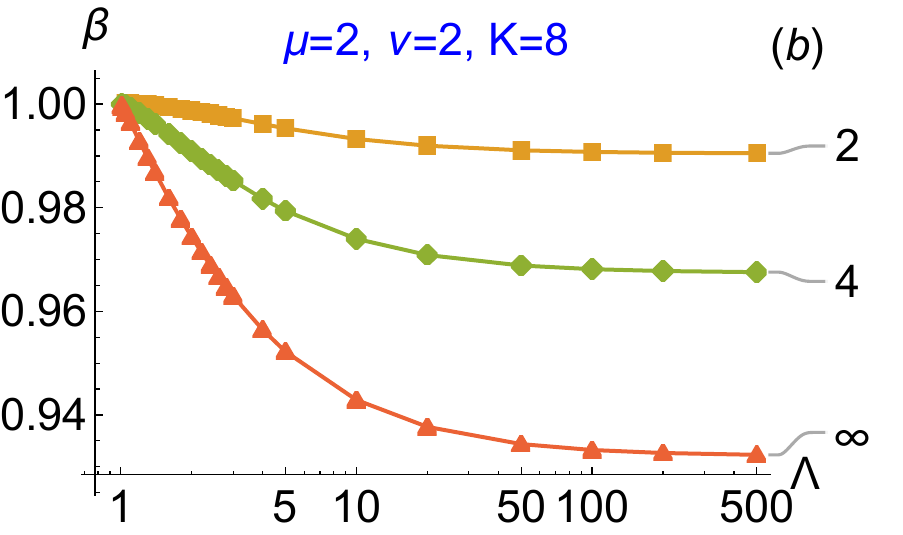}
  \\[1em]
  \includegraphics[width=0.49\linewidth]{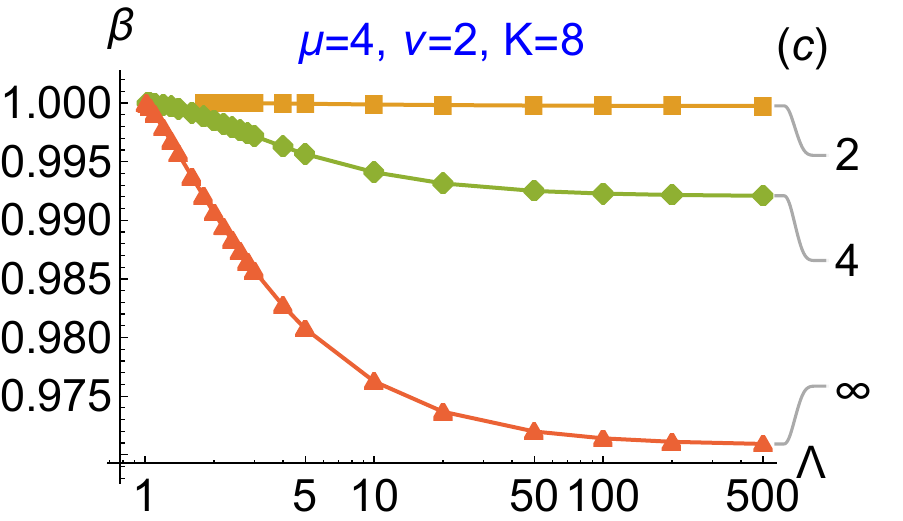}
  \hfill
  \includegraphics[width=0.49\linewidth]{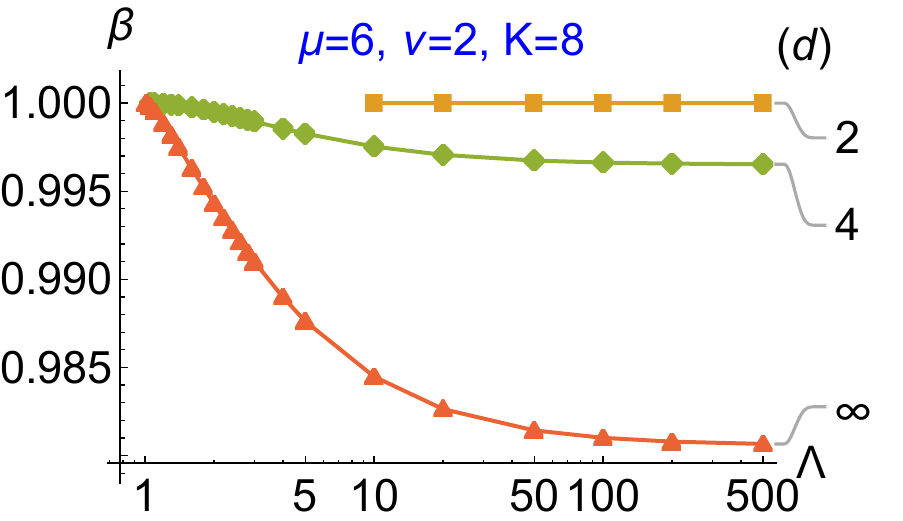}
  \caption{
  %
  Critical beta versus $\Lambda$ for the first model of the magnetic field  \eqref{4:31} at $K=8$, $\nu=2$ and index $\mu$ in the range from $\mu=1$ (picture a) to $\mu=6$ (picture d). The values of the parameter $k$ are shown to the right of the respective graphs.
  }\label{fig:GM-Isotropic_beta_vs_Lambda_K8}
\end{figure}

Figure \ref{fig:GM-Isotropic_beta_vs_Lambda_K8} shows plots of $\beta_{\text{crit}}$ versus $\Lambda$ for the first model of magnetic field \eqref{4:31} at $K=8$, $\nu=2$ and all tested values of $\mu$ and $k$. Comparison of figures \ref{fig:GM-Isotropic_beta_vs_Lambda_K8}(a)–(d) confirms the tendency noted in the previous section to increase critical value of beta as the magnetic mirrors steepen with increasing parameter $\mu$. We also see that  parabolic radial pressure profile ($k=1$) is unstable for all $\Lambda $ if $\mu\geq2$. For the next steepest profile ($k=2$), the stability zone disappears at $\mu=6$ and $\Lambda <10$, as shown in Fig.~\ref{fig:GM-Isotropic_beta_vs_Lambda_K8}(d).

\begin{figure*}
    \centering
  \includegraphics[width=0.3\linewidth]{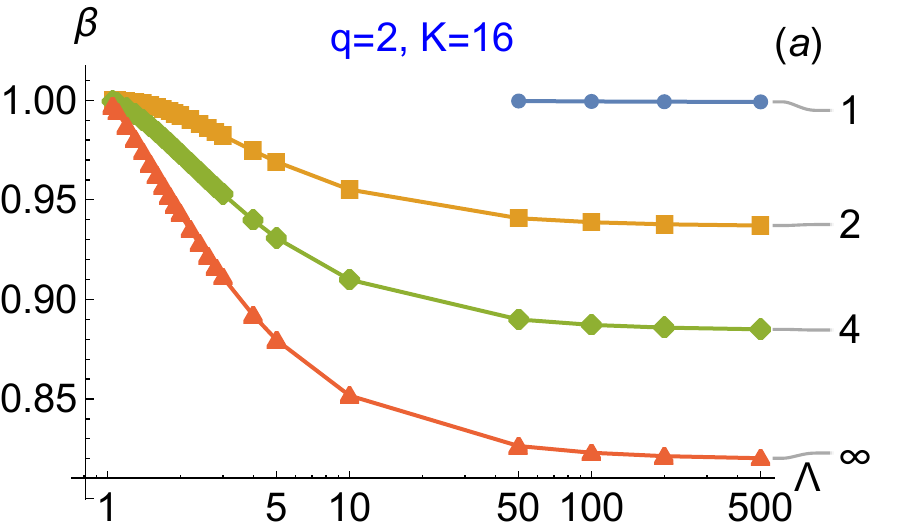}\hfil
  \includegraphics[width=0.3\linewidth]{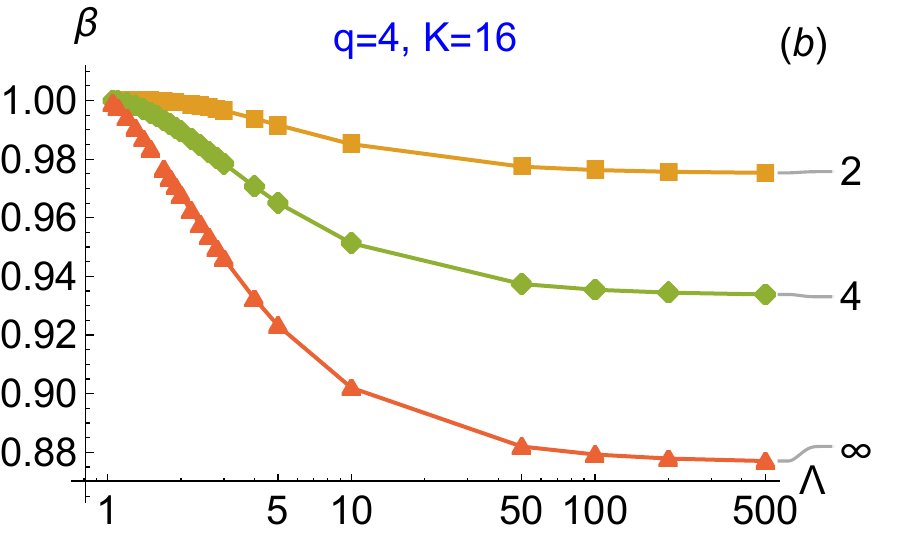}\hfil
  \includegraphics[width=0.3\linewidth]{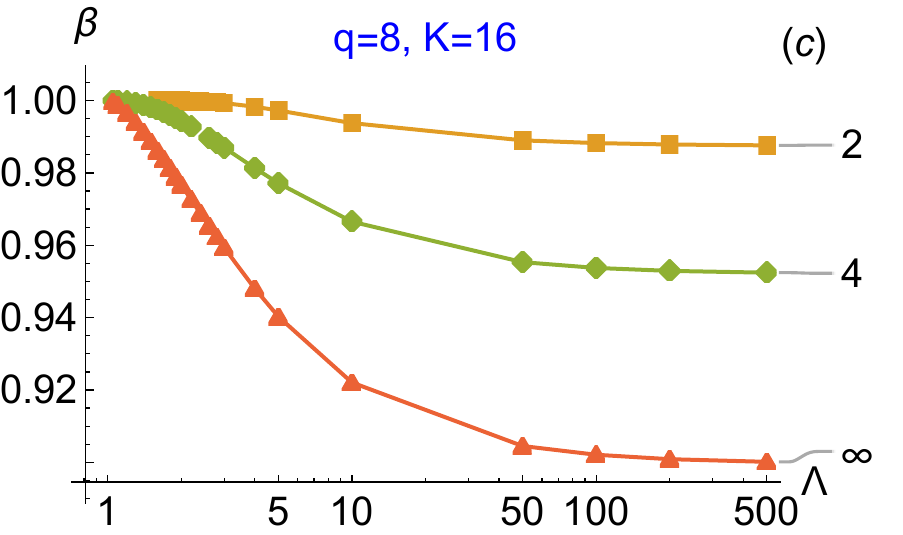}
  \caption{
    Critical beta versus $\Lambda$ for the second model of magnetic field  \eqref{4:35} at mirror ratio $K=16$ and index $q=\{2,4,8\}$. The values of parameter $k$  are shown to the right of the respective graphs.
  }
  \label{fig:GM-Kesner_beta_vs_Lambda}
\end{figure*}

Figure \ref{fig:GM-Kesner_beta_vs_Lambda} shows similar plots drawn for the second model of the magnetic field \eqref{4:35}. It illustrates the same tendency of increasing critical beta as the mirror throats become more narrow and more steep.


The main trends identified by our calculations are listed below:
\setlength{\itemindent}{0em}
\setlength{\leftmargini}{1em}
\begin{itemize}
  \item
  As expected, critical betas for the case $\Lambda=1$, when conducting lateral wall is removed  ($r_{w}/a=\infty $), have not been found.

  \item
  Next closest to $\Lambda=1$ checked value of $\Lambda$ was $1.01$ corresponding to very wide gap between plasma and conducting wall ($r_{w}/a=\num{14.1774}$). Unexpectedly, for such a wide gap, critical betas were found for some combinations of $k$, $\mu$, $\nu$, and $K$.

  \item
  As expected, the stability zone is the wider ($\beta_{\text{crit}}$ is smaller), the steeper the pressure profile is ($k$ is larger).

  \item
  As expected, the smaller the vacuum gap between the plasma and the conducting wall (the larger $\Lambda$), the wider the stability zone (the smaller $\beta_{\text{crit}}$).

  \item
  %
  %
  If the stability zone can in principle exist for a given set of parameters $K$, $\mu$, $\nu$, $q$ (i.e. if in the tables~\ref{tbl:K=20}–\ref{tbl:Kesner} for this set, the numerical value $\beta_{\text{crit}}$ is specified), then it occurs if the parameter $\Lambda$ exceeds some minimum value $\Lambda_{\min}$.
  This value is the smaller, the steeper the radial pressure profile (the larger the parameter $k$), the smoother the axial profile of the vacuum magnetic field (the smaller the parameters $\mu$ and $\nu$), and the smaller the mirror ratio $K$. The stability zone narrows with increasing $\Lambda_{\min}$ and may disappear altogether, first for smooth pressure profiles ($k=1, 2$), and then for steep ones ($k=4, \infty $).

  \item
  The first found critical value of beta for a given combination of parameters $k$, $\mu$, $\nu$, $q$, and $K$ is as close as possible to unity at $\Lambda=\Lambda_{\min}$. We did not set ourselves the goal of calculating $\Lambda_{\min}$ exactly, but simply chose the minimum values of $\Lambda$ from the available list of discrete values for which we calculated $\beta_{\text{crit}}$.

  \item
  %
  The stability zone is almost independent of the mirror ratio if $K\gtrapprox4$.

  \item
  The minimum $\beta_{\text{crit}}$ found for the studied set of radial and axial profiles is about $70\%$, which is slightly lower than the value $80\%$ reported in earlier publications.

\end{itemize}

\section{Wall stabilization combined with conductive ends}\label{s6}


Finally, it makes sense to perform calculations with the replacement of the boundary condition \eqref{3:12}, which describes the insulating ends of the trap, with the boundary condition \eqref{3:11}, which means that the plasma is frozen into the conductive end plates. The assumption of freezing into the ends is traditionally used in the theory of small-scale ballooning oscillations, but it has not been used before in the study of the hard ballooning mode.


Drawing an analogy with the works of \emph{D'Ippolito and Hafizi} \cite{DIppolitoHafizi1981PF_24_2274} and \emph{D'Ippolito and Myra} \cite{DIppolitoMyra1984PF_27_2256}, it could be expected that when the plasma is stabilized by simultaneously conducting ends and conducting side walls, two stability limits can exist. In particular, D'Ippolito and Myra investigated the stability of low-$m$ modes in an axisymmetric tandem mirror with an inverted (or hollow) step-wise pressure profile under the action of external ponderomotive rf-force. They found two critical beta values, $\beta_{\text{crit}1}$ and $\beta_{\text{crit}2}$: one at low beta due to the balancing of the ponderomotive force with the curvature drive, and one at high $\beta_{}$ due to the proximity of the conducting wall which enables magnetic line bending to balance the curvature drive. Corresponding to two critical values of beta, there are two zones of stability. The first zone exists at low plasma pressure, at $0<\beta<\beta_{\text{crit}1}$, and the second one exists at high pressure, at $\beta_{\text{crit}2}<\beta <$1. As calculations by D'Ippolito and Myra have shown, these two zones can merge.


%
To test our expectations, we examined the solution of Eq.~\eqref{3:01} with the boundary conditions \eqref{3:11} at $z=1$ and \eqref{5:01} at $z=0$. Preceding our calculations, we will show that under such boundary conditions there is indeed a stability zone at a small beta. To do this, we formulate a variational principle, that is, multiply Eq.~\eqref{3:01} by $\phi(z)$ and integrate the result over the interval from $z=-1$ to $z=+1$. After integrating by parts in the first line of Eq.~\eqref{3:01}, taking into account the boundary conditions, we obtain the integral equation
    \begin{multline}
    \label{6:01}
    \int_{-1}^{1}
    \left[
        \Lambda + 1
        -
        \frac{2\mean{\overline{p}}}{B_{v}^{2}}
    \right]
    \left(
        \der{\phi}{z}
    \right)^{2}
    \dif{z}
    =\\=
    \int_{-1}^{1}
    \phi^{2}
    \left[
        - \der{}{z}\left(
            \frac{B_{v}'}{B_{v}} + \frac{2a'}{a}
        \right)
        \left(
            1 - \frac{\mean{\overline{p}}}{B_{v}^{2}}
        \right)
    \right.
    \\
    \left.
    +
    \frac{\omega^{2}\mean{\rho}}{B_{v}^{2}}
    -
    \frac{
        2\mean{\overline{p}}
    }{B_{v}^{2}}\frac{a_{v}''}{a_{v}}
    \right.
    \\
    \left.
    -
    \frac{1}{2}\left(
            \frac{B_{v}'}{B_{v}} + \frac{2a'}{a}
    \right)^{2}
    \left(
        1 - \frac{\mean{\overline{p}}}{B_{v}^{2}}
    \right)
    \right]
    \dif{z}
    .
    \end{multline}
%
In the $\beta \to0$ limit, it becomes much simpler, since then $\mean{\overline{p}}=0$, $\left( {B_{v}'}/{B_{v}} + {2a '}/{a} \right)=0$, and we get the equality
    \begin{equation}
    \label{6:02}
    \int_{-1}^{1}
    \left[ \Lambda + 1 \right]
    \left( \der{\phi}{z} \right)^{2}
    \dif{z}
    =
    \omega^{2}
    \int_{-1}^{1}
    \left[
        \frac{\mean{\rho}}{B_{v}^{2}}
    \right]
    \phi^{2}
    \dif{z}
    .
    \end{equation}
%
Under the above boundary conditions, the derivative $\tder{\phi}{z}$ cannot be equal to zero identically over the entire integration interval, so both integrals on the left and right sides of this equality are greater than zero. Therefore, the square of the frequency is also positive, $\omega^{2}>0$, which means stability. Roughly estimating the derivative $\tder{\phi}{z}\sim -1$ (i.e., $\tder{\phi}{z}\sim -1/L$ in dimensional units), it is easy to see that $\omega $ corresponds to the frequency of Alfven oscillations, and as the gap between the plasma and the conducting wall decreases, it increases in proportion to $\sqrt{\Lambda}$.

\begin{figure*}
  \centering
  \includegraphics[width=0.3\linewidth]{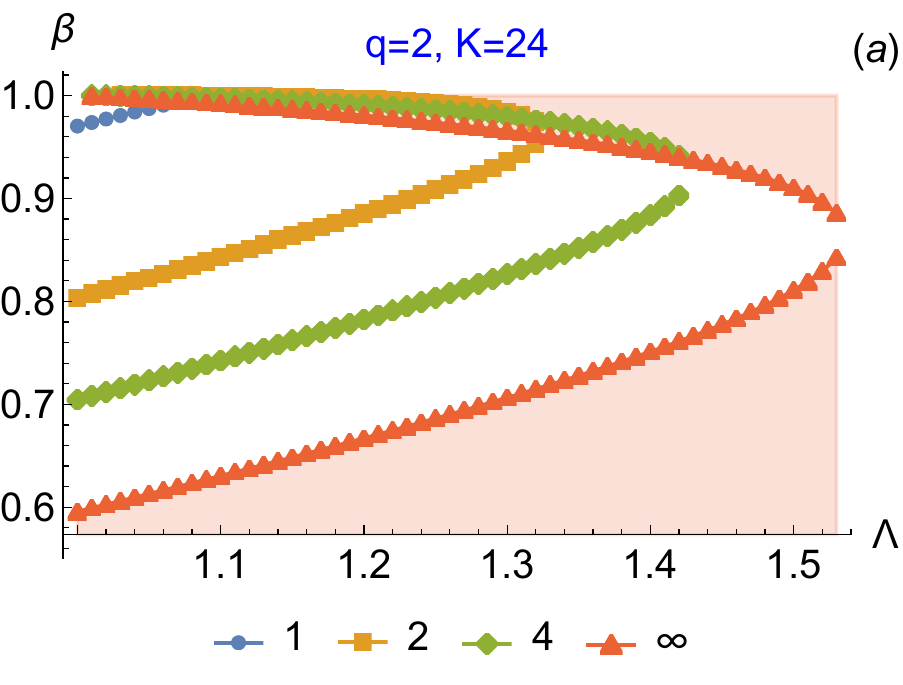}
  \includegraphics[width=0.3\linewidth]{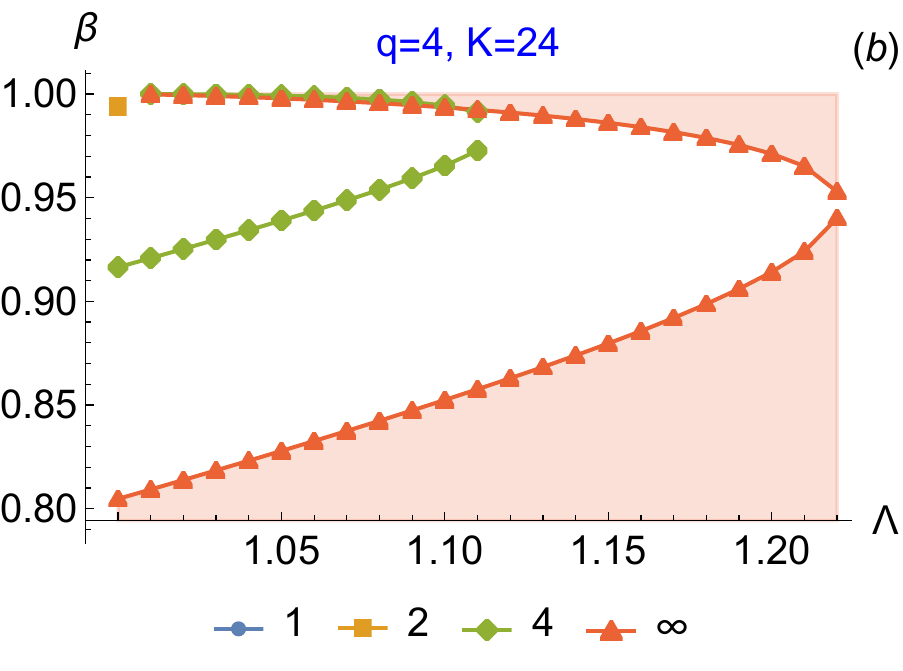}
  \includegraphics[width=0.3\linewidth]{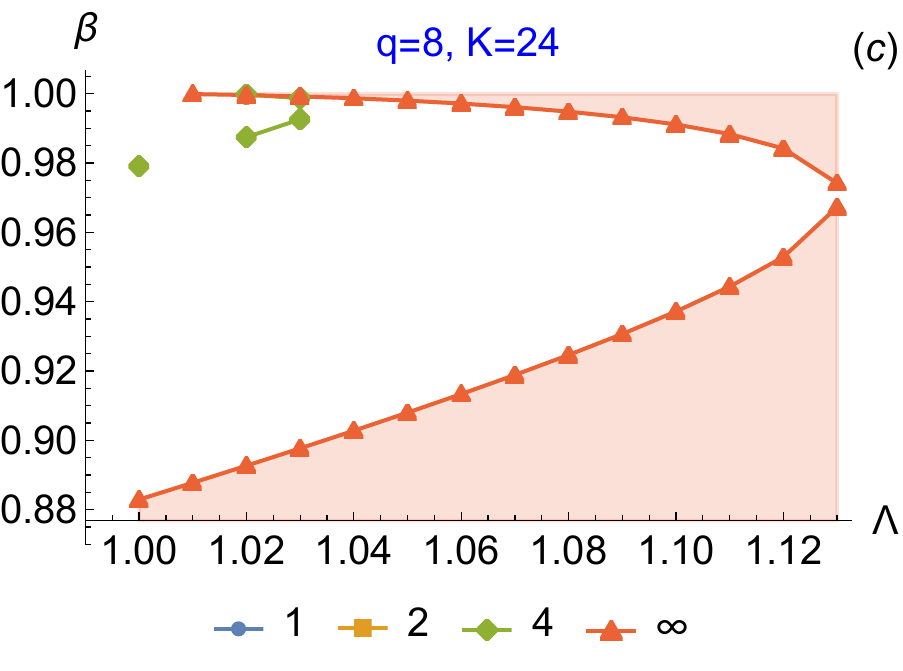}
  \\[0.5em]
  \includegraphics[width=0.3\linewidth]{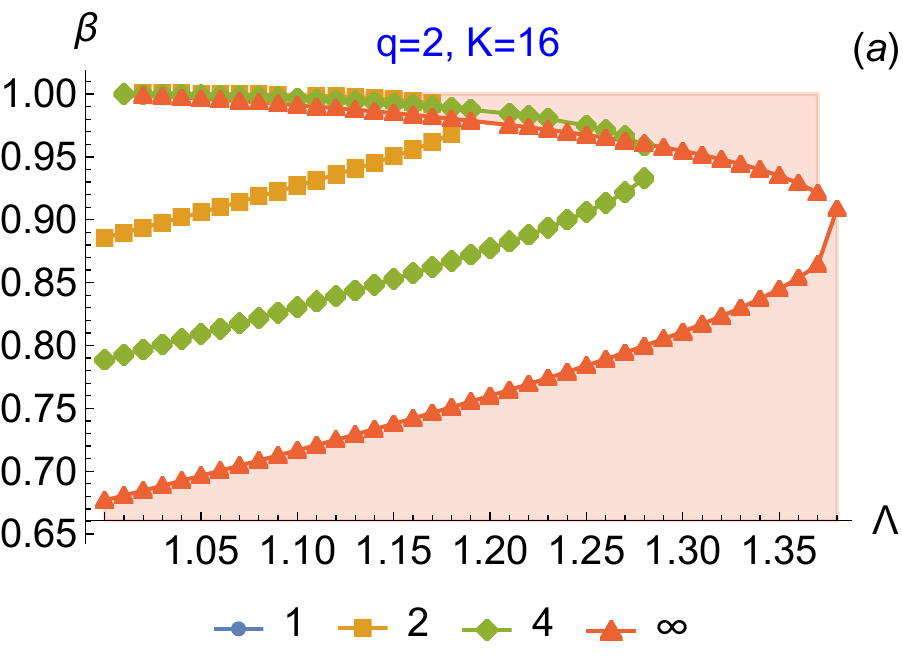}
  \includegraphics[width=0.3\linewidth]{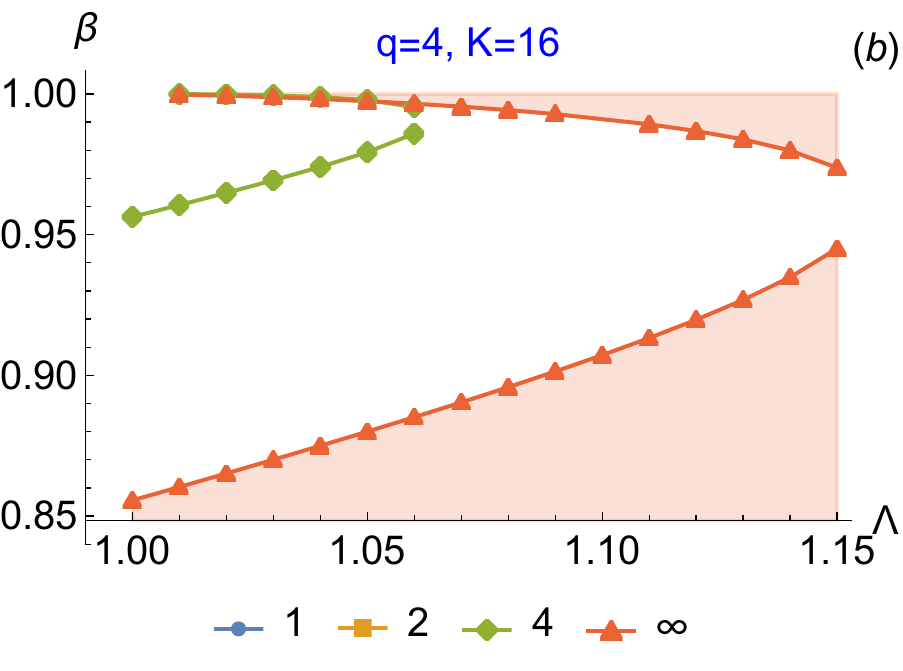}
  \includegraphics[width=0.3\linewidth]{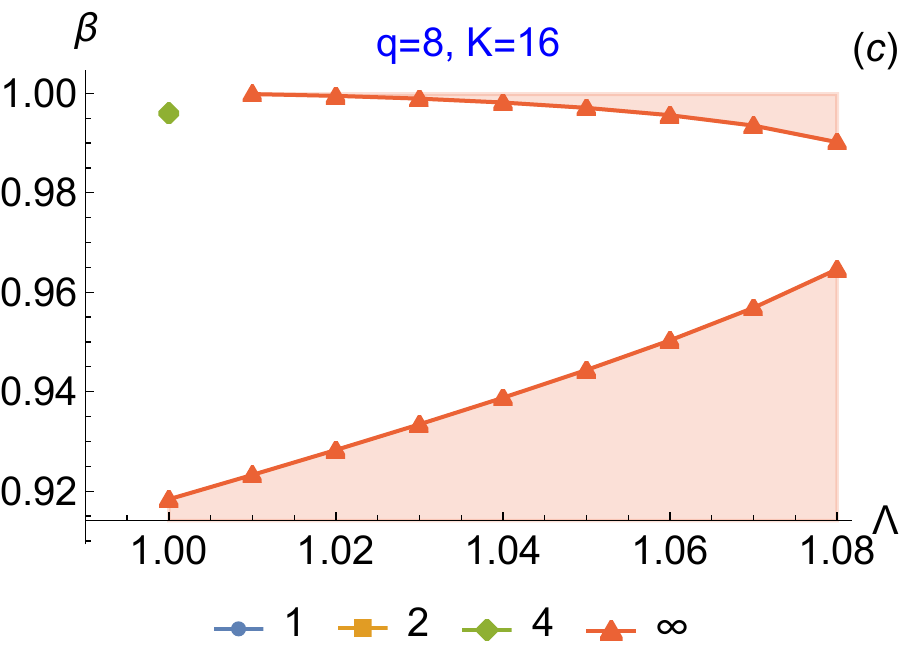}
  \\[0.5em]
  \includegraphics[width=0.3\linewidth]{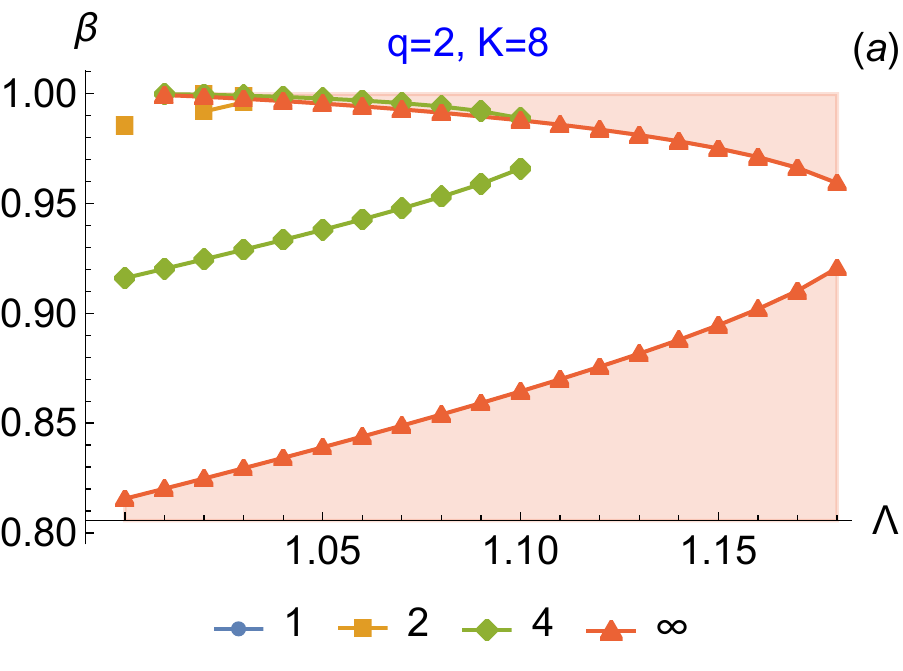}
  \includegraphics[width=0.3\linewidth]{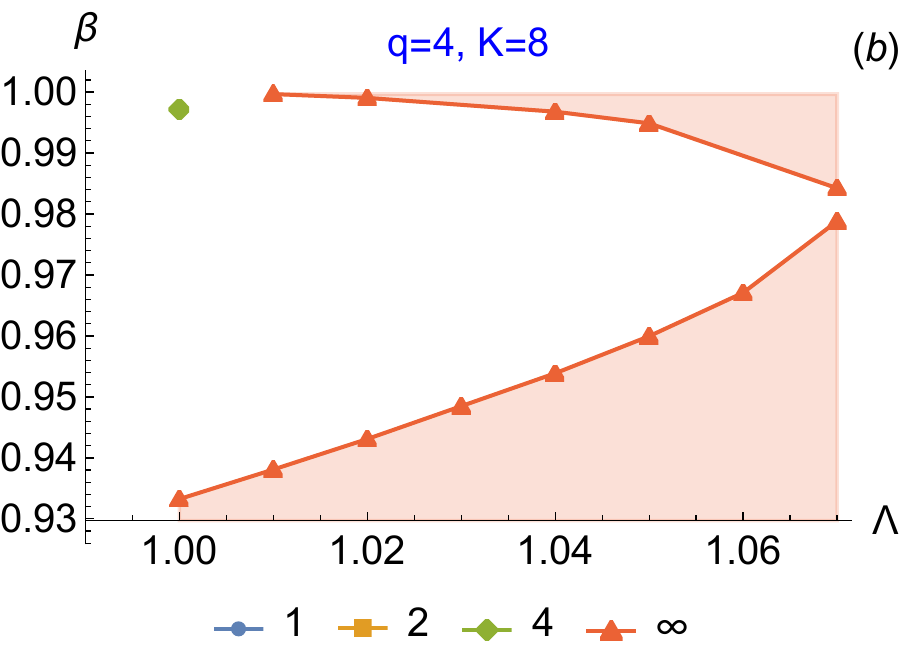}
  \includegraphics[width=0.3\linewidth]{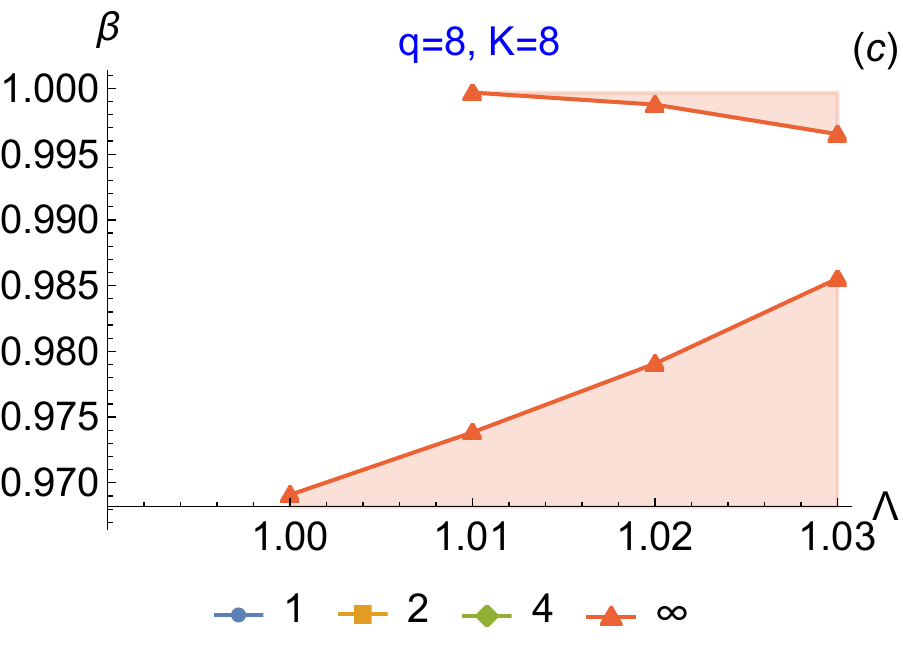}
  \caption{
    $\beta_{\text{crit}1}(\Lambda )$ (lower curve) and $\beta_{\text{crit}2}(\Lambda )$ (upper curve of the same color).
  }\label{fig:2022-GM-Kesner_beta_vs_Lambda}
\end{figure*}

To test our hypothesis, we performed a series of calculations for the second model of the vacuum magnetic field \eqref{4:35} for three values of the index $q=\{2,4,8\}$ and a mirror ratio from some set $K = \{24, 20, 16, 8, 4\}$. Figures \ref{fig:2022-GM-Kesner_beta_vs_Lambda} illustrate some of the results. Let's formulate the main observations:
\setlength{\itemindent}{0em}
\setlength{\leftmargini}{1em}
\begin{itemize}
  \item
    %
    In accordance with our expectations, two stability zones are found for moderate values of the $\Lambda$ parameter and a sufficiently large mirror ratio $K$. The lower zone $\beta < \beta_{\text{crit}1}$ exists even for $\Lambda =1$.

  \item
    Contrary to expectations, with other things being equal, the instability zone is maximum for the steepest pressure profile ($k=\infty $). Recall that in section \ref{s5} it is for such a profile that the instability zone had the minimum width.

  \item
    Contrary to expectations, it turned out that at a fixed $\Lambda $, the stability zones expand and can merge with a decrease in the mirror ratio $K$ and/or a decrease in the steepness of the radial pressure profile (a decrease in $k$).

  \item
    If an instability zone exists between two stability zones for some combinations of the parameters $k$, $q$, and $K$, then it disappears if $\Lambda > \Lambda_{\text{crit}}>1$. The value of $\Lambda_{\text{crit}}$ is the smaller, the smaller $k$, $K$ and the larger $q$. The largest value $\Lambda_{\text{crit}}=1.53$ ($r_{w}/a=\num{2.18485}$) in our calculations was found at $K=24$, $k=\infty $, $q=2$.

  \item
    For a smooth magnetic field profile with the index $q=2$, unstable zones are found for all investigated values of the mirror ratio $K = \{24, 20, 16, 8, 4\}$. For $q=4$ and $q=8$ unstable zones are found for $K=8$ and larger. For $K=4$ the unstable zone was found only for $q=2$ and radial profiles $k=4$, $k=\infty $.
\end{itemize}

When trying to repeat similar calculations for the first model of the magnetic field \eqref{4:31}, our Wolfram  \emph{Mathematica}$^{\copyright}$ code warned of possible problems with the $pf$ interpolation function. We got rid of such diagnostic messages by smoothing function  \eqref{4:31} near the point $z=0$ as described in Appendix \ref{A1}. These calculations showed that the smoothing of the magnetic field profile near the median plane (i.e.\ increasing $\delta $ in Eq.~\eqref{A1:03}) effectively contributes to the reduction of the instability zone. Implementing alternative smoothing method Eq.~\eqref{A1:04}, we also concluded that the smoothing of the field near magnetic mirrors is not as significant as near midplane.

Summing up everything said in this section, we can state the following:
\setlength{\itemindent}{0em}
\setlength{\leftmargini}{1em}
\begin{itemize}
  \item
    With a not too large mirror ratio, a sufficiently steep magnetic field, and a sufficiently smooth pressure profile, the rigid ballooning mode $m=1$ can be stabilized at any value of beta even if the side conducting wall is dismantled.

    \item
    If, on the other hand, the side wall is close enough to the plasma, then, in combination with end conductors, this mode can be stabilized for any axial magnetic field profile and any radial pressure profile.

    \item
    Smoothing the magnetic field profile at the center of the trap is an effective way to reduce and even eliminate the instability zone.

\end{itemize}

\section{Conclusions}\label{s9}

In the present work, we have studied the wall stabilization of the $m=1$ rigid ballooning mode in an open axially symmetric trap using a conducting cylindrical wall of the chamber surrounding the plasma column. To simplify the problem, we used the isotropic plasma approximation, planning in the next article to present the results of calculations for an anisotropic plasma. In contrast to the works of predecessors, who studied only the not quite realistic case of a plasma with a sharp-boundary radial profile, we considered four variants of a diffuse pressure profile with different degrees of steepness, specified by the index $k$, as well as many variants of the axial profile of the vacuum magnetic field given by the functions \eqref{4:31} and \eqref{4:35} with different values of the indices $\mu$, $\nu$, $q$ at different mirror ratios $K$.

Stabilization by a conducting wall becomes possible if the plasma beta (that is, the dimensionless ratio of the plasma pressure to the magnetic field pressure) exceeds a certain critical value $\beta_{\text{crit}}$. Therefore, our goal was to calculate this critical value and study its dependence on the radial pressure profile, the axial profile of the magnetic field, the mirror ratio, and the magnitude of the vacuum gap between the plasma and the conducting wall. For calculations, we developed a numerical code in the Wolfram \emph{Mathematica}$^{\copyright}$ system, which solved the equation \eqref{3:01}, previously derived by Linda LoDestro, using the shooting method.

On the whole, our calculations confirmed the assertion available in the literature that for wall stabilization of an isotropic plasma, the beta parameter must exceed $80\%$. However, we have found examples of plasma configurations with a critical beta of $70\%$. Investigating the dependence of $\beta_{\text{crit}}$ on the parameters of the problem, we found that the mirror ratio has a relatively weak effect on the value of $\beta_{\text{crit}}$. The dependence of $\beta_{\text{crit}}$ on the parameters $k$ is more significant (the larger $k$, the steeper the radial profile, the smaller $\beta_{\text{crit}}$), $\mu$, $\nu$ and $q$ (the larger $\mu$, $\nu$ and $q$, the shorter the magnetic mirrors, the larger $\beta_{\text{crit}}$), as well as on the parameter $\Lambda $ (the larger $\Lambda $, the smaller the gap between the plasma and the conducting wall, the smaller $\beta_{\text{crit}}$). The stability zone can formally exist even at very wide vacuum gap between conducting lateral wall and the plasma surface (when $\Lambda \to 1$), although the width of such a zone tends to zero, since $\beta_{\text{crit} } \to 1$.

We also studied the stabilization of the rigid ballooning mode by a combination of the conductive lateral wall and conductive end plates, which simulate the attachment of MHD end stabilizers to the central cell of a mirror trap. Our calculations have shown the great efficiency of this method of stabilization. We found the presence of two zones of stability. The low beta zone is due to the curvature drive being balanced by the end plate effect, and the upper beta zone is due to the curvature drive being compensated by the proximity of the conductive lateral wall. These two zones merge, making the entire range of allowable values of beta $0<\beta<1$ stable as the mirror ratio decreases or the vacuum gap between the plasma and the side wall decreases.

The key feature of an isotropic plasma is the constancy of pressure along the magnetic field lines. In an anisotropic plasma, the pressure depends on the magnitude of the magnetic field on the field line, usually decreasing towards the magnetic mirrors. This fact noticeably complicates the calculations, since the equation \eqref{3:04}, generally speaking, cannot be solved with respect to $B$ as simply as in the case of an isotropic plasma. In the next article, we will present the results of calculations for a fairly realistic dependence of $p_{\bot}$ on $B$ when the equation \eqref{3:04} is solved without using numerical methods.

Another continuation of studies on the stabilization of the rigid ballooning mode can be the rejection of the simplifying assumption $\Lambda=\const$. The constancy of $\Lambda $ in the equation \eqref{3:01} implies that the radius $r_{w}(z)$ of the conducting wall depends in a complex way on the plasma radius $a(z)$ (and hence on $\beta_{}$) from the radial pressure profile. A more realistic case is $r_{w}=\const$. In addition,
special profiling of a conductive wall might expand the stability zone to some extent by lowering the value of $\beta_{\text{crit}}$.
%
%
In particular, it is necessary to check the proposal to stabilize the ballooning instability in the “diamagnetic Beklemishev bubble” (figure \ref{fig:GM-a_vs_z}(c)) using conical MHD stabilizers, which are proposed to be installed at the edges of the bubble, where the curvature of the magnetic field lines is maximal.

\begin{acknowledgements}


    This work has been done in the framework of ALIANCE collaboration \cite{Bagryansky+2020NuclFusion_60_036005}. It was supported by Chinese Academy of Sciences President’s International Fellowship Initiative (PIFI) under the Grants No.~2022VMA0007, No.~2022VMB0001, No.~2021VMB0013, and Chinese Academy of Sciences International Partnership Program under the Grant No.~116134KYSB20200001.


\end{acknowledgements}

\section*{ORCID iDs}


\noindent
Igor KOTELNIKOV \href{https://orcid.org/0000-0002-5509-3174}{https://orcid.org/0000-0002-5509-3174}
\\
Vadim PRIKHODKO \href{https://orcid.org/0000-0003-0199-3035}{https://orcid.org/0000-0003-0199-3035}
\\
Dmitri YAKOVLEV \href{https://orcid.org/0000-0002-2224-4618}{https://orcid.org/0000-0002-2224-4618}
\\
Qiusun ZENG \href{https://orcid.org/0000-0001-9572-2206}{https://orcid.org/0000-0001-9572-2206}

\appendix
\section{Regularization of magnetic field model}\label{A1}

%
The first model of the magnetic field has the peculiarity that the derivative $B_{v}'$ of functions \eqref{4:31} does not vanish at the ends of the integration interval $z=\pm1$, as was assumed when deriving the equation \eqref{4:11}. Physically, this should mean that magnetic coils of such a small size are installed in the mirror throats that, on the scale under consideration, the coils can be considered "point". At the size of such a coil, the derivative $B_{v}'(\pm1)$ quickly vanishes. Accordingly, the second derivative $B_{v}''(\pm1)$ of any function \eqref{4:31} must be supplemented with the delta function $\delta(z\mp1)$. We write this rule in symbolic form:
    \begin{equation}
    \label{A1:32}
    B_{v}''(z)
    \Rightarrow
    B_{v}''
    - B_{v}'(1)\,\delta(z-1)
    + B_{v}'(-1)\,\delta(z+1)
    .
    \end{equation}
%
In addition, it should be noted that the function \eqref{4:31} for $\mu=1$ has a break at $z=0$, since the derivative
    \begin{equation}
    \label{A1:33}
    |z|' = -1 + 2\theta(z).
    \end{equation}
experiences a jump there. Second derivative
    \begin{equation}
    \label{A1:34}
    |z|'' = 2\delta(z)
    \end{equation}
%
enters Eq.~\eqref{4:11} under the integral sign through the vacuum curvature $a_{v}''$ and also contains the delta function $\delta(z)$. If $\mu>1$, this delta function enters the integrand, being multiplied by $|z|^{\mu-1}$, so it makes a zero contribution to the integral, but in the case of $\mu=1$ calculating the integral needs special care. Note that the combination $\mu=1$, $\nu=2$ is remarkable in that the function corresponding to it minimizes the absolute value of the integral in the Rosenbluth-Longmire
 criterion \cite{RosenbluthLongmire1957AnnPhys_1_120}, which determines the stability condition for flute oscillations in open traps (see \cite{Kotelnikov2021V2e}).

Another, more natural in terms of physics, way of dealing with magnetic field singularities in the first model is to smooth these singularities. One variant of smoothing was achieved by replacing
    \begin{equation}
    \label{A1:03}
    |z|
    \Rightarrow
    \frac{\sqrt{z^2+\delta^2}-\delta}{\sqrt{1+\delta^2}-\delta}
    ,
    \end{equation}
where it was assumed that the parameter $\delta$ is sufficiently small; in particular, the values $\delta =\{0.1, 0.05, 0.01\}$ were tested in Section \ref{s6}, where the combined effect of wall and end conductors is examined. These calculations showed that the smoothing of the magnetic field profile near the median plane (i.e.\ increasing $\delta $) effectively contributes to the reduction of the instability zone.

As a result of the replacement \eqref{A1:03}, the derivative $B_{v}'$ vanishes smoothly as $z\to0$, but not as $z\to\pm1$. Therefore, a second replacement was also tested
    \begin{equation}
    \label{A1:04}
    |z|
    \Rightarrow
    \frac{1}{2} \left[
        \text{sn}((2 z-1) \K(n)|n)+1
    \right]
    ,
    \end{equation}
which smoothes the derivative $B_{v}'$ both for $z\to0$ and $z\to\pm1$. This replacement includes Jacobi elliptic function $\text{sn}$ and complete elliptic integral of the first kind $\K(n)$. The parameter $n$ can vary widely, but suitable values lie in the range $-2<n<-5$. With $n=-4$, our calculations in Section \ref{s6} gave approximately the same results as those obtained for the replacement \eqref{A1:03} with $\delta =0.05$. From this fact we concluded that the smoothing of the field near magnetic mirrors is not as significant as near midplane.

\section{Transfer of boundary conditions}\label{A2}


%
When choosing the first model of the magnetic field \eqref{4:31}, the coefficients of the LoDestro equation \eqref{3:01} can have singularities at the end points of the interval $0\leq z \leq 1$ in the form of delta functions $\delta(z)$ and $\delta(z-1)$. To get around the problem of singularities, it is enough to find a solution to Eq.~\eqref{3:01} on the interval $0+<z<1-$, from which the endpoints $z=0$ and $z=1$ are excluded. Excluding these points from the domain of the solution, we must set the boundary conditions at the point $z=0+$ a little to the right of the point $z=0$ and at the point $z=1-$ a little to the left of the point $z=1$.

Recall that in Section \ref{s5} we used the boundary conditions $\phi'(0)=0$ and $\phi'(1)=0$. However, due to the presence of delta functions in the coefficients of Eq.~\eqref{3:01}, the derivatives $\phi'(0+)$ and $\phi'(1-)$ will no longer be equal to zero.
%
%
From a mathematical point of view, to calculate the increment of the derivative $\phi'(0+)-\phi'(0)$, one should integrate Eq.~\eqref{3:01} over an infinitesimal interval from $z=0$ to $z= 0+$. Similarly, to calculate the difference $\phi'(1)-\phi'(1-)$, one must repeat the integration from $z=1-$ to $z=1$. In the Wolfram \emph{Mathematica}$^{\copyright}$, there is actually no need to calculate integrals.
The values of the derivatives $\phi'(0+)=BC0$ and $\phi'(1-)=BC1$ can be calculated using the built-in utility \texttt{Coefficient}. It easily extracts the coefficient in front of $\delta(z)$ on the right-hand side of Eq.~\eqref{3:01}. All terms in this coefficient come from the second pair of square brackets. Let's denote this coefficient as $N0$, and the coefficient at $\phi''$ (that is, the sum of terms inside the first pair of square brackets) as $D0$. Both coefficients must be evaluated at $z=0$. The desired value of the derivative $\phi'(0+)=BC0$ is found by the formula $BC0=-N0/2D0$. Since $N0$ contains the factor $\phi(0)$, this boundary condition links $\phi'(0+)$ and $\phi(0)=\phi(0+)$.


In a similar way, one can find the boundary condition $\phi'(1-)=BC1$ on the right boundary if, before substituting function $B_{v}(z)$ from Eq.~\eqref{4:31} into Eq.~\eqref{3:01}, make the substitution $|z|\to 1-|1-z|$. Then some terms appear in the second square bracket that contain the delta function $\delta(z-1)$. The coefficient in front of this delta function will be denoted as $N1$, and the coefficient of $\phi''$ will be denoted as $D1$. Both these coefficients should be taken at $z=1$. The desired value of the derivative $\phi'(1-)=BC1$ is found by the formula $BC1=N1/2D1$. Since $N1$ contains the factor $\phi(1)$, this boundary condition links $\phi'(1-)$ and $\phi(1)=\phi(1-)$.


Passing Eq.~\eqref{3:01} to the \texttt{Parametric\-NDSolve\-Value} utility, we excluded delta function from its right-hand side. In the Wolfram \emph{Mathematica}$^{\copyright}$ this is done by the rule \text{/.\{DiracDelta[z\_] -> 0\}}. In addition, we substituted $\omega^{2}=0$ because we wanted to calculate the critical (marginal) value of beta, and not the ballooning frequency for a given beta.

%
The boundary conditions for the \texttt{Parametric\-NDSolve\-Value} utility were specified on the left boundary as
    \begin{equation}
    \label{A2:01}
    \phi(0)=1,
    \qquad
    \phi'(0)=BC0
    \end{equation}
instead of Eq.~\eqref{5:01}. As to Eq.~\eqref{5:02}, it should be substituted with
    \begin{equation}
   \label{A2:02}
    pf[\beta,\Lambda]'[1]=BC1
    .
    \end{equation}


%

\end{document}